\newif\ifpreprint

\preprintfalse 

\ifpreprint
\documentclass[journal=jctcce,manuscript=letter]{achemso}
\else
\documentclass[journal=jctcce,manuscript=letter,layout=twocolumn]{achemso}
\fi

\usepackage[T1]{fontenc} 

\usepackage{amsmath}
\usepackage{newtxtext,newtxmath}

\usepackage{graphicx}
\usepackage{dcolumn}
\usepackage{braket}
\usepackage{multirow}
\usepackage{threeparttable}
\usepackage{xspace}
\usepackage{verbatim}
\usepackage[version=4]{mhchem} 
\usepackage{comment}
\usepackage{color,soul}

\usepackage{mathtools}
\usepackage[dvipsnames]{xcolor}
\usepackage{xspace}
\usepackage{ifthen}

\usepackage{qcircuit}

\usepackage{graphicx,longtable,dcolumn,mhchem}
\usepackage{rotating,color}
\usepackage{lscape}
\usepackage{amsmath}
\usepackage{dsfont}
\usepackage{soul}
\usepackage{physics}
\newcolumntype{d}{D{.}{.}{-1}}

\newcommand{\AVTZ}{\emph{aug}-cc-pVTZ}

\newcommand{\AVFZ}{\emph{aug}-cc-pV5Z}

\usepackage[colorlinks = true,
            linkcolor = blue,
            urlcolor  = black,
            citecolor = blue,
            anchorcolor = black]{hyperref}

\definecolor{goodorange}{RGB}{225,125,0}
\definecolor{goodgreen}{RGB}{5,130,5}
\definecolor{goodred}{RGB}{220,50,25}
\definecolor{goodblue}{RGB}{30,144,255}

\newcommand{\note}[2]{
\ifthenelse{\equal{#1}{F}}{
\colorbox{goodorange}{\textcolor{white}{\footnotesize \fontfamily{phv}\selectfont #1}}
    \textcolor{goodorange}{{\footnotesize \fontfamily{phv}\selectfont #2}}\xspace
}{}
\ifthenelse{\equal{#1}{R}}{
\colorbox{goodred}{\textcolor{white}{\footnotesize \fontfamily{phv}\selectfont #1}}
    \textcolor{goodred}{{\footnotesize \fontfamily{phv}\selectfont #2}}\xspace
}{}
\ifthenelse{\equal{#1}{N}}{
\colorbox{goodgreen}{\textcolor{white}{\footnotesize \fontfamily{phv}\selectfont #1}}
    \textcolor{goodgreen}{{\footnotesize \fontfamily{phv}\selectfont #2}}\xspace
}{}
\ifthenelse{\equal{#1}{M}}{
\colorbox{goodblue}{\textcolor{white}{\footnotesize \fontfamily{phv}\selectfont #1}}
    \textcolor{goodblue}{{\footnotesize \fontfamily{phv}\selectfont #2}}\xspace
}{}
}

\usepackage{titlesec}

\usepackage[fontsize=11pt]{scrextend}
\titleformat{\section}
{\normalfont\sffamily\bfseries\color{Blue}}
{\thesection.}{0.25em}{\uppercase}

\titleformat{\subsection}
{\normalfont\sffamily\bfseries}
{\thesubsection}{0.25em}{}

\titleformat{\subsubsection}
{\normalfont\sffamily}
{\thesubsubsection}{0.25em}{}

\titleformat{\suppinfo}
{\normalfont\sffamily\bfseries}
{\thesubsection}{0.25em}{}

\titlespacing*{\section}{0pt}{0.5\baselineskip}{0.01\baselineskip}
\titlespacing*{\subsection}{0pt}{0.125\baselineskip}{0.01\baselineskip}
\titlespacing*{\subsubsection}{0pt}{0.125\baselineskip}{0.01\baselineskip}

\author{Rudraditya Sarkar}
	\affiliation[CEISAM, Nantes]{Universit\'e de Nantes, CNRS,  CEISAM UMR 6230, F-44000 Nantes, France}
\author{Martial Boggio-Pasqua}
	\affiliation[LCPQ, Toulouse]{Laboratoire de Chimie et Physique Quantiques, Universit\'e de Toulouse, CNRS, UPS, France}
\author{Pierre-Fran\c{c}ois Loos}
	\email{loos@irsamc.ups-tlse.fr}
	\affiliation[LCPQ, Toulouse]{Laboratoire de Chimie et Physique Quantiques, Universit\'e de Toulouse, CNRS, UPS, France}
\author{Denis Jacquemin}
	\email{Denis.Jacquemin@univ-nantes.fr}
	\affiliation[CEISAM, Nantes]{Universit\'e de Nantes, CNRS,  CEISAM UMR 6230, F-44000 Nantes, France}
		
\setkeys{acs}{usetitle=true}

\let\oldmaketitle\maketitle
\let\maketitle\relax
     \title{Benchmarking TD-DFT and Wave Function Methods for Oscillator Strengths and Excited-State Dipole Moments}
\date{\today}

\begin{tocentry}
	\centering
	\includegraphics[width=.50\textwidth]{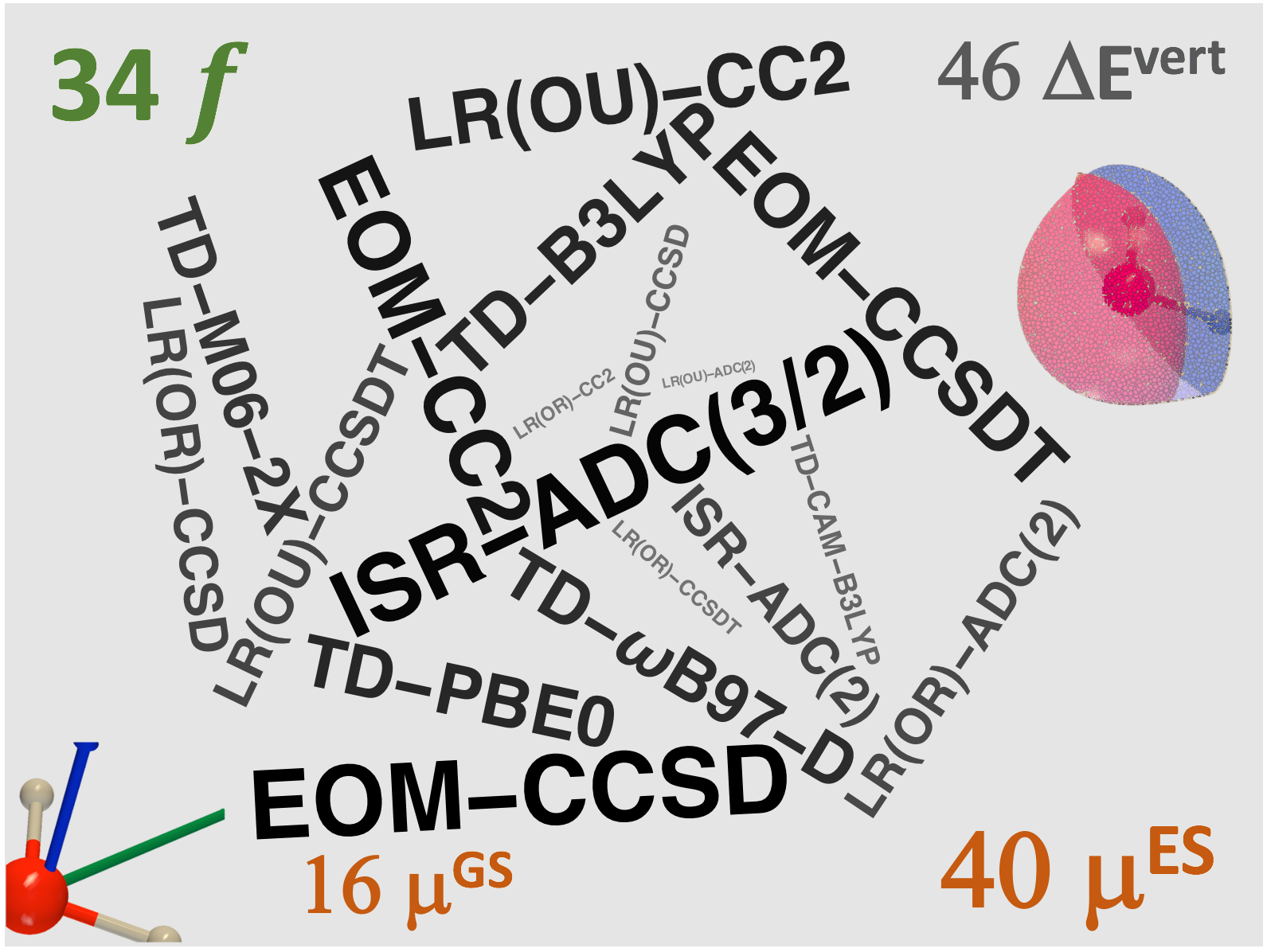}
\end{tocentry}

\begin{document}

\ifpreprint
\else
\twocolumn[
\begin{@twocolumnfalse}
\fi
\oldmaketitle

\begin{abstract}
Using a set of oscillator strengths and excited-state dipole moments of near full configuration interaction (FCI) quality determined for small compounds, we benchmark the performances of several single-reference wave function methods
(CC2, CCSD, CC3, CCSDT, ADC(2), and ADC(3/2)) and time-dependent density-functional theory (TD-DFT) with various functionals (B3LYP, PBE0, M06-2X, CAM-B3LYP, and $\omega$B97X-D). 
We consider the impact of various gauges (length, velocity, and mixed) and formalisms: equation of motion (EOM) \emph{vs} linear response (LR),  relaxed \emph{vs} unrelaxed orbitals, etc. 
Beyond the expected accuracy improvements and a neat decrease of formalism sensitivy when using higher-order wave function methods, the present contribution shows that, for both ADC(2) and CC2,
the choice of gauge impacts more significantly the magnitude of the oscillator strengths than the choice of formalism, and that CCSD yields a notable improvement on this transition property
as compared to CC2. For the excited-state dipole moments, switching on orbital relaxation appreciably improves the accuracy of both ADC(2) and CC2, but has a rather small effect at the CCSD level.
Going from ground to excited states, the typical errors on dipole moments for a given method tend to roughly triple. Interestingly, the ADC(3/2) oscillator strengths and dipoles are significantly more
accurate than their ADC(2) counterparts, whereas the two models do deliver rather similar absolute errors for transition energies. Concerning TD-DFT, one finds: i) a rather negligible impact of the 
gauge on oscillator strengths for all tested functionals (except for M06-2X); ii) deviations of ca.~0.10 D on ground-state dipoles for all  functionals; iii) strong differences between excited-state 
dipoles obtained with, on the one hand, B3LYP and PBE0, and on the other hand, M06-2X, CAM-B3LYP, and $\omega$B97X-D, the latter group being markedly more accurate; iv) the better overall performance 
of CAM-B3LYP for the two considered excited-state properties.  Finally, for all investigated properties, both the accuracy and consistency obtained with the second-order wave function approaches, ADC(2) and 
CC2, do not clearly outperform those of TD-DFT, hinting that assessing the accuracy of the latter (or selecting a specific functional) on the basis of the results of the former is not systematically a well-settled strategy.
\end{abstract}

\ifpreprint
\else
\end{@twocolumnfalse}
]
\fi

\ifpreprint
\else
\small
\fi

\noindent

\section{Introduction}

The accurate modeling of electronically excited states (ESs) remains one of the main goals pursued by theoretical and computational chemists. \cite{Gon12,Dre05,Ada13a,Lau13,Roo07,Kry06,Pie02,Sne12,Gho18,Bla20,Loo20c} Indeed, theory is often required to 
support or complement  photophysical and photochemical experiments, whose results are typically analyzed on the basis of both empirical models and first-principles calculations.  Along the years, thanks to the fantastic development and implementation 
efforts of many research groups around the world, several \emph{ab initio} methods have become applicable to chemically-relevant problems, and one can highlight the well-established time-dependent density-functional theory (TD-DFT) formalism, \cite{Run84,Cas95,Ada13a} 
the second-order algebraic diagrammatic construction [ADC(2)] scheme for the polarization propagator, \cite{Sch82,Tro99,Har14,Dre15} the second-order approximate coupled-cluster (CC2) method, \cite{Chr95} as well as the emerging Bethe Salpeter equation 
(BSE/$GW$) formalism, \cite{Bet51,Str88,Bla18,Bla20} and the similarity-transformed equation-of-motion coupled-cluster with singles and doubles (STEOM-CCSD) method \cite{Noo99,Dut18} as black-box single-reference methods able to deliver ES energies and 
properties for rather large compounds in complex environments.  However, none of these five methodologies is able to yield chemically accurate excitation energies, i.e., average absolute errors smaller than 1 kcal.mol$^{-1}$ (or 0.043 eV).\cite{Loo20c} Additionally, 
the computationally most efficient approach, namely TD-DFT, is plagued by a significant dependency on the selected exchange-correlation functional (XCF) in its traditional adiabatic formulation. \cite{Lau13} 

Another noteworthy issue is that theoretical calculations must be both accurate enough and deliver data allowing meaningful comparisons with measurements. 
This is a significant difficulty, as the most directly accessible theoretical quantities, i.e., the vertical excitation energies (VEEs), have no direct experimental equivalents. Therefore, the sets of theoretical best estimates (TBEs) of 
VEEs proposed by Thiel and coworkers, \cite{Sch08,Sau09,Sil10c} and by some of us, \cite{Loo18a,Loo19c,Loo20a,Loo20d} have been mostly used to benchmark lower-order methods but do not pave the way to straightforward 
theory-experiment comparisons. To lift this difficulty, the obvious choice is to focus on the so-called 0-0 energies, \cite{Die04b,Sen11b,Loo19a,Loo19b} but this requires computing ES vibrations, which is computationally expensive. In addition,  comparisons between 
measured and predicted 0-0 energies is often limited to the lowest ES and still provides an almost purely energetic probe. Indded, it has been evidenced that the accuracy of the selected structures and vibrations is rather irrelevant in 
the final theoretical estimate. \cite{Sen11b,Loo19a} 

Even for ground states (GSs), it has been argued that such energetic metric provides only a rather incomplete assessment of the \emph{pros} and \emph{cons} of any specific methods. For example, a Kohn-Sham DFT energy can be accurate even if the
underlying density is inaccurate. \cite{Med17} As stated by Hait and Head-Gordon, molecular dipoles are, despite their intrinsic limitations, handy and valuable quantities for probing the quality of the density (or wave function). \cite{Hai18,Hai20} In an ES
framework, one may use not only the ES dipole moments ($\mu^\mathrm{ES}$) but also the oscillator strengths ($f$) or other transition-related quantities to estimate the quality of ES wave functions or densities.  Advantageously, both $f$ and $\mu^\mathrm{ES}$ can be measured
experimentally, although the accuracy of such measurements clearly depends on both the nature of the ES and the experimental technique, as we discussed in the introduction of an earlier work. \cite{Chr21} 

To allow well grounded benchmarks of ES properties, we have recently developed a set of theoretical $\mu^\mathrm{ES}$ and $f$ values of (near) full configuration interaction (FCI) quality for a significant set of small molecules and ESs. \cite{Chr21}  
To attain near-FCI quality, we systematically increased the excitation degree of the coupled-cluster (CC) expansion up to quintuples (i.e., CCSDTQP) using a series of increasingly large correlation-consistent atomic basis sets including diffuse functions 
so as to be as close as possible from both radial and angular near-completeness. \cite{Gin19} The goal of the present work is to use these reference data (as well as new additional values computed for the present study) to benchmark a large set of wave function and 
density-based approaches for $f$ and $\mu^\mathrm{ES}$.

In assessing theoretical models for these two ES properties, an additional complexity comes from the fact that various ``options'' are available to compute these, which is not the case for VEEs. First, the actual value of the oscillator strength 
naturally depends on the gauge of the interaction operator between the quantum system and the applied periodic field, and these quantities can be determined in the so-called length, velocity, or mixed gauges, the former being the typical default in 
most electronic structure software packages.  The results obtained with these three gauges are equivalent only when the wave function is exact (or in the complete basis set limit for approximate methods, such as the random phase approximation or TD-DFT, \cite{Dre05} which fulfill the Thomas-Reiche-Kuhn sum rule, \cite{Tho25,Rei25,Kuh25} the fulfillment of this sum rule being commonly used as a test of the degree of completeness of the one-electron basis set). \cite{Sau19} Taking neon, the nitrogen molecule, and water as examples, 
Pawloski \emph{et al.}~demonstrated that the gauge invariance improves significantly going from CC2 and CCSD to CC3. \cite{Paw04} This is, to the very best of our knowledge, the only study tackling this issue in a systematic way, the sole exception 
being an earlier work limited to CCSD by Pedersen and Koch. \cite{Ped98}  

Second and most importantly, there exist two different ways for calculating molecular properties, which are, in general, not equivalent for approximate wave functions. One of the possible approaches consists in computing a given property as a derivative of the energy with respect to 
the perturbation strength.  In such a case, the variation of the wave function parameters are typically obtained through a Lagrangian formalism that provides a rigorous mathematical framework for such purposes. \cite{Hel89,Koc90,Chr95,Hat03,Hat05c} The well-known linear-response 
(LR) \cite{Mon77,Koc90,Chr98d,Kal04} formalism of CC theory follows this philosophy, and such formalisms can be applied for ADC as well. \cite{Hat05c}  In LR, one can additionally choose to perform the calculations within the ``orbital-relaxed'' (OR) or 
``orbital-unrelaxed'' (OU) scheme, the latter neglecting the orbital response due to the external perturbation (e.g., the electric field for the dipole moments). \cite{Sal87,Tru88} 
In other words, at the LR(OU)-CC level, only the variation of the CC amplitudes with respect to the external perturbation is taken into account.
Again, OU and OR do lead to the same results only when the underlying function is exact. Alternatively, one can compute molecular properties directly from the expectation value of the corresponding operator for the physical observable.
This second route is followed in CC calculations performed in the well-known equation-of-motion (EOM) \cite{Row68,Sta93} formalism, as well as in ADC calculations done within the so-called intermediate state representation (ISR). \cite{Sch91,Sch10,Dre15}
Indeed, propagator methods such as ADC are not intrinsically designed to compute excited-state wave functions and properties, and the ISR has been introduced to palliate this. In terms of orbital relaxation, the ISR formalism can be viewed as an intermediate between OR and OU, \cite{Hod19c}
whereas  the main theoretical distinction between LR and EOM is that the more expensive LR formalism takes into account the relaxation of the ground-state CC amplitudes due to the external perturbation (hence providing size-intensive transition properties), while the cheaper EOM 
approach freezes them during the computation of the perturbation but includes the contribution of the reference determinant to the transition properties. \cite{Sta93,Koc94} It is noteworthy, that all these formalisms [LR(OR), LR(OU), EOM, or ISR] systematically provide the same VEEs 
irrespective of the truncation order of the CC or ADC expansion. Yet, they deliver distinct oscillator strengths and dipole moments from one another (except again in the case of  the exact wave function) as the level of wave function relaxation (at the correlated level) differs. 
In terms of numerical experiments, several groups have provided estimates: i) Caricato \emph{et al.}~studied the oscillator strengths obtained at the LR-CCSD and EOM-CCSD levels, \cite{Car09b} and concluded that significant differences only occur for large systems;
ii) Kannar and Szalay also compared EOM and LR oscillator strengths obtained using CC2 and CCSD (see below); \cite{Kan14} (but no comparisons between EOM-CC3 and LR-CC3 has, to the best of our knowledge, been made available to date); 
iii) the Dreuw group has also evidenced  for several examples that the differences between various frameworks tend to become insignificant as the expansion order of the ADC or CC series is increased. \cite{Hod19c} Finally, neither the derivative-based formalism nor the 
expectation-value-based formalism can be considered superior in general.  

In short, there is a plethora of ``options''  making the benchmarking of ES properties clearly more involved than that of VEEs. Several valuable benchmarks of oscillator strengths and excited-state dipole moments are, of course, available for various test sets, 
and we provide an overview of these works below.

\section{Literature survey}


\subsection{Oscillator strengths}

Let us start with oscillator strengths. These have been the subject of several assessments, most of them being focussed on TD-DFT's reliability for such quantities. \cite{Taw04,Miu07,Tim08,Sil08,Car10d,Sza12,Kan14,Sau15,Jac16b,Rob18} As early as 2007, Champagne 
and coworkers compared TD-DFT oscillator strengths obtained with several XCFs in a series of substituted benzenes to both experimental and CC2/CCSD values. \cite{Miu07} They concluded that a large share of exact exchange is required to obtain 
accurate results and recommended the use of BH\&HLYP. In their seminal 2008 paper, the Thiel group compared LR-CC2,  LR-CCSD and CASPT2 oscillator strengths determined with the TZVP basis set for their well-known set of compounds. \cite{Sil08} In contrast 
to the VEEs, they did not define TBEs, but highlighted the high degree of correlation between LR-CC2 and LR-CCSD values, whereas the CASPT2 oscillator strengths were found to be generally larger than their CC counterparts. They, next, extended this 
comparison to several TD-DFT  approaches, \cite{Sil08} and reported mean absolute errors (MAEs) of 0.128 (0.075), 0.113 (0.055), 0.096 (0.044) and 0.069 (0.062) for BP86, B3LYP, BH\&HLYP, and DFT/MRCI, respectively when using their own CASPT2 
(LR-CC2) TZVP values as references. In 2010, the same group investigated basis set effects on the LR-CC2 oscillator strengths. \cite{Sil10b} The same year, Caricato and coworkers compared $f$ values for eleven compact organic compounds (69 ESs) obtained with EOM-CCSD and
TD-DFT, \cite{Car10d} both methods being applied with a very extended atomic basis set. Using the former approach as benchmark, these authors evidenced that CAM-B3LYP was the best performer amongst the 28 tested XCFs. 
In 2014, Dreuw's group performed an extensive comparison of oscillator strengths obtained with ADC(2) and ADC(3/2) (within the ISR formalism), LR-CC2, LR-CCSD, as well as literature data for the molecules of the Thiel's set. They concluded that 
ADC(3/2) delivers very accurate oscillator strengths. \cite{Har14} The same year, Kannar and Szalay determined LR-CC3/TZVP oscillator strengths for Thiel's set, and even LR-CCSDT/TZVP values for a subset of 15 ESs. \cite{Kan14} They tested both the LR and 
EOM formalisms for both CC2 and CCSD and found that LR slightly outperforms EOM, whereas the CCSD oscillator strengths are significantly more accurate than their CC2 equivalents. This is likely the most advanced benchmark of 
oscillator strengths available to date (for wave function methods). The data produced in Ref.~\citenum{Kan14} were subsequently employed to benchmark both SOPPA \cite{Sau15} and BSE/$GW$ oscillator strengths, \cite{Jac16b} the former showing significantly larger MAEs than the latter. 
In Ref.~\citenum{Jac16b}, ADC(2), CC2 and BSE/$GW$ oscillator strengths determined for large organic dyes are reported and their relative accuracy discussed. Yet again, such comparisons do not rely on indisputable references.
We note that all the studies on oscillator strengths mentioned above are performed in the length gauge.

\subsection{Excited-state dipole moments}

As for oscillator strengths, one can find several benchmark studies dealing with excited-state dipoles, most of them aiming at finding the most suitable XCF in a TD-DFT context. \cite{Fur02,Kin08,Sil08,Won09b,Tap09,Gui10,Sil10,Hel11,Jac16d,Hod19c,Hod20} 
The first investigation in this vein is likely due to Furche and Ahlrichs who considered ten $\mu^\mathrm{ES}$ in tiny compounds and compared the performances of five XCFs (BLYP, BP86, PBE, B3LYP, and PBE0) to experimental values. \cite{Fur02} 
Quite surprisingly, they found that the errors are larger with the global hybrids (B3LYP and PBE0) than with the three GGAs (BLYP, BP86, and PBE), the latter delivering MAEs in the range 0.11--0.12 D.  In 2008, King compared TD-DFT excited-state dipoles determined 
with two functionals to the corresponding CC values obtained by EOM-CCSD and LR-CC3 values for 29 ESs in pyrrole and furan. He reported respective MAEs around 1.1 D and 0.5 D for these two compounds when considering the B97-1 functional. \cite{Kin08} 
The same year, Thiel and coworkers reported MAEs of 0.75, 0.59, and 0.61 D with BP86, B3LYP, and BH\&HLYP, using their CASPT2/TZVP dipoles as reference. \cite{Sch08} Again, they later computed LR-CC2/{\AVTZ} values for ES dipole moments. \cite{Sil10b}  
The selection of OU or OR for the computation of $\mu^\mathrm{ES}$ is apparently not specified in that latter work. In 2011, Hellweg compared GS and ES dipole moments computed with the OR approach combined with LR-ADC(2), LR-CC2, and several spin-scaled 
variants to experimental values for ten molecules. \cite{Hel11} He reported a MAE of ca.~0.2 D for both of these second-order models, highlighting that they are reasonably accurate, yet far from flawless. In 2016, one of us compared the LR(OR)-CC2/{\AVTZ} and 
TD-DFT/{\AVTZ} (16 XCFs) excess dipole moments (i.e., $\Delta\mu = \mu^\mathrm{ES} - \mu^\mathrm{GS}$) determined in 30 organic dyes, \cite{Jac16d} which led to MAEs around 1 D for hybrid XCFs and 1.5 D for semi-local XCFs. Finally, in the recent and very 
detailed work from Ref.~\citenum{Hod19c}, Hodecker \textit{et al.}~compared, in particular, the Lagrangian and ISR formalisms for several ADC variants [ADC(1), ADC(2), ADC(2)-x, ADC(3/2), and ADC(3)] considering both the GS and ES dipoles. They found that the 
ISR ADC(2) $\mu^\mathrm{ES}$ values are rather close from their LR(OU) analogs. In addition, they also showed that, within the LR framework, the differences between OR and OU GS dipole moments are slightly larger with CCSDT than with CC3, and that the impact of 
orbital relaxation is larger for $\mu^\mathrm{ES}$ than for $\mu^\mathrm{GS}$, irrespective of the CC excitation level. The same group very recently assessed the performance of unitary CC theory for several ES dipole moments in water, hydrogen fluoride, 
4-cyanoindole, and 2,3-benzofuran. \cite{Hod20} For the two former (latter) molecules FCI/3-21G (experimental) values were considered as reference.

As evidenced by the above literature survey, previous benchmark studies typically rely on significantly less accurate reference values  than the present contribution. Moreover, they are generally focussed on a subset of specific approaches and/or molecules. 
Here, we have specifically designed the present study to be as general as possible (yet limited to small-sized compounds), while reporting comparisons between various formalisms for a given wave function method, as well as further assessments against highly-accurate TBEs.

\section{Methods}

\subsection{Molecules, geometries and basis sets}

\begin{figure}[htp]
\centering
 \includegraphics[width=.75\linewidth]{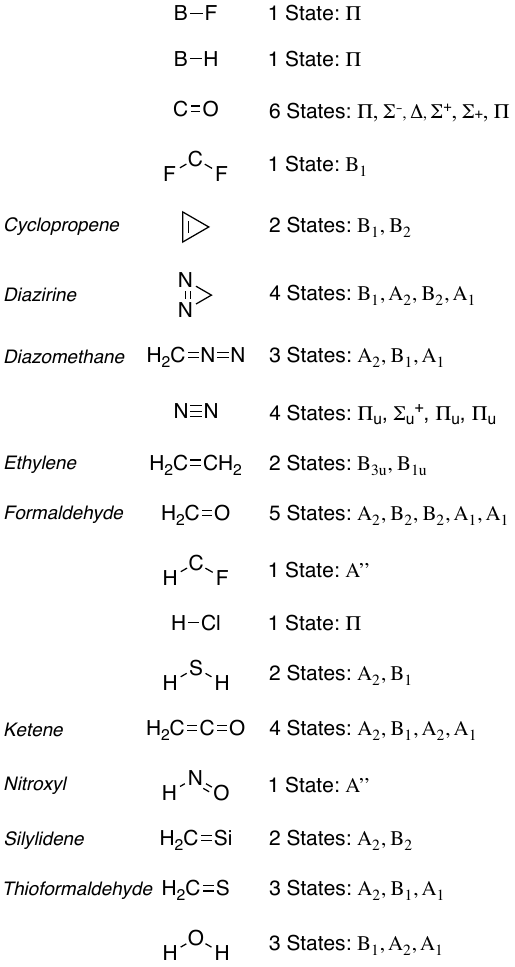}
  \caption{Representation of the molecules constituting the present benchmark set together with the list of considered ESs (in increasing energy order for each molecule).  
  The interested reader is referred to our previous works \cite{Loo18a,Chr21} and the SI of the present article for more details regarding the relative energies and nature of these ESs.}
   \label{Figure-1}
\end{figure}

The molecules and states considered in the present study are represented in Fig.~\ref{Figure-1}. The corresponding geometries have been obtained at the CC3/{\AVTZ} level of theory \cite{Chr95b,Koc95} and are given in the {Supporting Information (SI)}. Note that several
structures come from previous works, \cite{Loo18a,Loo20d,Chr21} whereas a few additional optimizations have been specifically performed for the present study with DALTON 2017 \cite{dalton} and CFOUR 2.1 \cite{cfour} applying default parameters in both cases.

All our computations of VEEs, $f$,  $\mu^\mathrm{GS}$, and $\mu^\mathrm{ES}$ have been performed with the {\AVTZ} basis set while enforcing the frozen-core (FC) approximation during the ground- and excited-state calculations.
It is well known that energies and properties derived from density-based and wave function-based methods exhibit rather different basis set dependencies. \cite{Hat12,Loo19,Gin19} However, this augmented triple-$\zeta$ basis set was chosen because it stands as the \emph{de facto} standard in ES calculations.
We refer the interested reader to Section S6 of the SI where one can find a basis set study for the water molecule.  In addition, our previous work also contains many data regarding  basis set effects at various CC levels.\cite{Chr21}

\subsection{Reference values}

For the smallest molecules, our reference values have been taken from Ref.~\citenum{Chr21}, choosing the TBE/{\AVTZ} values listed in this earlier work. We have added to this original data set, a set of five molecules (CF$_2$,
cyclopropene, diazirine, diazomethane, and ketene) encompassing an additional 14 ESs. To this end, we used exactly the same protocol as in our earlier work.  \cite{Chr21} Briefly, we have applied the MRCC (2017 and 2019) program \cite{Kal20,mrcc}  
to perform LR-CC calculations for $f$ (length gauge), $\mu^{\mathrm{GS}}$, and $\mu^{\mathrm{ES}}$ with CCSD, \cite{Pur82,Scu87,Koc90b,Sta93,Sta93b} CCSDT, \cite{Nog87,Scu88,Kuc01,Kow01,Kow01b} and CCSDTQ, \cite{Kuc91,Kal03,Kal04,Hir04} 
combined with several atomic basis sets of increasingly large size. The OR variant of the LR formalism was systematically applied for the CC calculations. Next, we used the traditional incremental approach, \cite{Kal04,Bal06,Kam06b,Wat12,Fel14,Fra19} consisting 
in estimating the LR-CCSTQ contribution determined with a small basis set to correct the LR-CCSDT/{\AVTZ} values, so as to obtain TBE values.  In our previous work, \cite{Chr21} it was shown that the difference between  LR-CCSTQ and  LR-CCSTQP 
dipoles and oscillator strengths are negligible. At this stage, we would like to stress that all our reference data include (at least) CCSDTQ corrections and have been determined with the most ``ambitious'' scheme, e.g., within the  LR(OR) formalism.
The interested reader is referred to Tables S1--S5 of the SI where one can find details regarding these new calculations as well as comparisons with existing experimental and theoretical literature data. For three of the molecules, we also performed
additional CASPT2/{\AVTZ} calculations. \cite{And90,And92}  The CASPT2 calculations were performed on top of the state-averaged (SA) complete active space self consistent field (CASSCF) wave function. The SA-CASSCF wave functions have been
obtained by considering the ground state and at least the excited state of interest. The information regarding the construction of the SA-CASSCF wave functions can be found in the relevant Tables in the SI. We tackled the intruder state problem in CASPT2 both by 
increasing the size of active spaces as well as by introducing a level shift parameter (0.3 a.u.). \cite{Roo95}  One set of CASPT2 calculations are performed by introducing the ionization-potential-electron-affinity (IPEA) shift (0.25 a.u.). \cite{Gio04} 
 All the SA-CASSCF and CASPT2 (rs2 contraction level) calculations have been performed with MOLPRO 2019 program. \cite{molpro,MolPro19}.

\subsection{Computational details}

We used a large panel of codes to perform our calculations.  We generally applied default parameters for all programs, as comparisons between the results obtained with different codes showed totally negligible differences (see Ref. \citenum{Chr21} and the SI).
Nonetheless, in the specific case of ES CC calculations with Q-CHEM 5.3, we had to significantly tighten the default settings to obtain numerically accurate dipole moments. As stated above, we systematically employed the {\AVTZ} basis set and enforce the 
FC approximation during all  wave function calculations. TD-DFT calculations are likewise performed in the FC approximation.  For the ADC and CC calculations performed with TURBOMOLE 6.4 as well as the ADC calculations performed with Q-CHEM 5.3, the RI 
approach was applied with the default auxiliary basis sets associated with {\AVTZ}.

Because ADC is a polarization propagator-based method, one does not have directly access to the ground-state wave function because such approach aims at a direct expansion of the response functions in orders of the electron fluctuation potential without 
reference to a specific ground-state wave function. \cite{Hat05c,Hod19c} However, the natural and computationally convenient choice is to choose MP2 and MP3 as ground-state methods for ADC(2) and ADC(3), respectively, which is furthermore consistent in 
terms of expansion order in the fluctuation potential.

\subsubsection{Ground-state dipoles.} 

The computation of $\mu^\mathrm{GS}$ with wave function approaches have been performed with the following codes: MRCC (2017 and 2019)  \cite{Kal20,mrcc} for CCSDT and CCSD (both OR and OU), 
DALTON \cite{dalton} for CC3 (OU), CFOUR for CC3 (OR), TURBOMOLE \cite{Bal20,Turbomole} for CC2 and MP2 (both OR and OU), and Q-CHEM 5.3 \cite{Kry13} for MP2 (ISR). At the DFT level, the 
dipoles have been computed with GAUSSIAN 16, \cite{Gaussian16} selecting five popular hybrid functionals, namely B3LYP, \cite{Bec93,Fri94} PBE0, \cite{Ada99,Erz99} M06-2X, \cite{Zha08b} CAM-B3LYP, 
\cite{Yan04} and $\omega$B9X-D.\cite{Cha08b}  The default \emph{ultrafine} grid was used for all DFT and TD-DFT calculations. For the wave function approaches many consistency checks have been 
performed between various codes, including MRCC, Q-CHEM, GAUSSIAN, CFOUR, and DALTON (see Table S8 in the SI), and, as expected, only trifling differences (0.001 D) could be detected between 
various implementations.

\subsubsection{Oscillator strengths.} 

Our LR-CC2, LR-CCSD, and LR-CC3 oscillator strengths have been obtained in the three gauges thanks to DALTON. The EOM-CC2, EOM-CCSD, and EOM-CC3 values in length gauge were obtained with 
$e^T$ 1.0, \cite{eT} whereas the EOM-CCSD values in the two other gauges were obtained with GAUSSIAN 16. The LR-CCSDT values (length gauge) were determined with the MRCC program. Q-CHEM was 
used to obtain ADC(2) and ADC(3/2) oscillator strengths in the ISR formalism (length gauge), whereas LR-ADC(2) values were obtained with TURBOMOLE in the three gauges. Finally, all TD-DFT calculations 
were realized with GAUSSIAN, applying the same XCFs as above. 

\subsubsection{Excited-state dipoles.}

For the $\mu^\mathrm{ES}$ values, we selected MRCC again for LR-CCSDT and LR-CCSD (both OR and OU), TURBOMOLE for LR-CC2 and LR-ADC(2) (both OU and OR),  CFOUR for EOM-CC2 and EOM-CCSD,
Q-CHEM for the ISR-ADC(2) and ISR-ADC(3/2), and GAUSSIAN for all TD-DFT calculations. Again, the interested reader is referred to Table S16 in the SI for additional consistency checks between the various codes,
including PSI-4 1.2\cite{Psi4} for the EOM-CC dipoles. 

As a final remark, let us stress that we provide, for molecules having a single non-zero dipole component, the signs of dipole moments in the Tables of the SI, so that a sign change from the GS to an ES is clear.  
However, we do not account for signs while computing MSEs, Max($+$) and Max($-$), only its magnitude. In such a way, a negative MSE indicates an underestimated dipole amplitude rather than a specific orientation.

\section{Results and Discussion}

\subsection{On the challenge and consequences of state mixing}

For most of the ESs treated herein, the identification of the considered states with all tested theoretical approaches could be easily achieved using the usual criteria for such task (energy, symmetry, nature and symmetry of the involved 
molecular orbitals, oscillator strength, sign and magnitude of the dipole moment). Therefore, except for a few cases, the assignments are unambiguous. Nevertheless, it should be highlighted that, in contrast to benchmarks focussed on VEEs for
which ambiguity in a specific assignment is not systematically a dramatic issue, state mixing is often a major problem when assessing $f$ and $\mu^{\mathrm{ES}}$.  Indeed, the mixing of two ESs of the same symmetry can strongly affect their 
properties while their relative energies remain similar, making assignments less settled. 
Let us discuss a few challenging cases.

For the three $\Pi_u$ ESs of dinitrogen, that are close in energy, the computed $f$ values are very strongly basis set and method dependent, \cite{Chr21} so that several conflicting analyses of their nature can be found in the 
literature. \cite{Odd85,Ben90,Neu04} We have here considered the symmetry of the dominant molecular orbital pair in our assignment, \cite{Chr21} but it is quite obvious that the relative magnitudes of the oscillator strengths
cannot be definitively determined: for the second $\Pi_u$ ES, our TBE is $0.015$, but values one order of magnitude larger are obtained with CC2, ADC(2) and all tested XCFs in a TD-DFT context.

For the third ES of diazirine, a Rydberg transition of $B_2$ symmetry (see Table S3 in the SI), our TBE value is $7.44$ eV with $\mu^{\mathrm{ES}}=3.03$ D. Many methods exhibit significant state mixing, an effect particularly
pronounced with the two range-separated hybrids that yield incorrect $\mu^{\mathrm{ES}}$ values ($0.89$ D with CAM-B3LYP and $-0.50$ D with $\omega$B97X-D), although the computed vertical energies are within $0.3$ eV of the TBE
value (see Table S6). For the Rydberg ES of cyclopropene ($B_1$), one also observes significant state mixing at the TD-DFT level, although it is less dramatic than for diazirine.

The highest $A_1$ transition of formaldehyde considered here has a $\pi \rightarrow \pi^\star$ nature (TBE values: $\Delta E_{\mathrm{vert}} = 9.44$ eV, $f = 0.13$, and $\mu^{\mathrm{ES}} = 1.30$ D). Although the CC2 energy is reasonable
($9.55$ eV), the CC2 oscillator strength is markedly too small ($0.05$--$0.06$) and the ES dipole has the incorrect direction ($-2.59$/$-2.98$ D). These large errors are likely related to significant state 
mixing with a slightly higher-lying $A_1$ ES at $9.91$ eV that is built up on the same molecular orbital pairs, which also shows a significant oscillator strength ($0.05$--$0.06$), but still an incorrectly oriented dipole ($-1.17$/$-1.68$ D).

For the corresponding $\pi \rightarrow \pi^\star$ $A_1$ ES in thioformaldehyde, it has been shown that the contributions of the quadruples in the CC framework are rather large. \cite{Chr21} Additionally, TD-B3LYP
yields another close-lying ES that significantly mixes with the $A_1$ state, making the B3LYP $f$ value of $0.10$ artificially much smaller than the one computed with the four other XCFs ($0.19$--$0.21$), yet closer from the TBE ($0.14$). 
The TD-B3LYP $\mu^{\mathrm{ES}}$ ($7.41$ D) is totally off target (TBE of $1.18$ D), while all other XCFs do provide much better estimates. Selecting the other close-lying TD-B3LYP ES would also yield
a very inaccurate value of $\mu^{\mathrm{ES}}$ ($-4.17$ D), so that state mixing yields incorrect dipoles for both ESs.

When comparing several gauges and formalisms (e.g., length \textit{vs} velocity or EOM \emph{vs} LR), the above-described difficulties are rather irrelevant; one still compares clearly equivalent states. Therefore, we have not discarded 
these challenging situations during such comparisons, nor when discussing VEEs. The issue is, of course, also irrelevant for $\mu^{\mathrm{GS}}$. However, when assessing the relative performances of methods with respect to the TBEs for
$f$ and  $\mu^{\mathrm{ES}}$, one is left with two reasonable choices: i) including these difficult states in the statistics on the basis that state mixing is inherent to the quality of the assessed method (the exact treatment 
would not suffer from this particular deficiency),  a choice of course resulting in large average deviations, or ii) discarding these problematic ESs as there is no clear one-to-one correspondence between methods. This latter option comes 
at the cost of removing the most difficult cases for a specific method, hence making the results look better than they truly are. In the body of the text, we went for the second
solution and we have discarded accordingly the following ESs in the performance evaluation of the various models for both $f$ and $\mu^{\mathrm{ES}}$: i) the three $\Pi_u$ ESs of dinitrogen, ii) the $B_2$ transition in diazirine, and iii) the 
$A_1$ $\pi \rightarrow \pi^\star$ of both formaldehyde and thioformaldehyde. Of course, one could also find other ``borderline'' cases as there is no definitive answer to state identification  when benchmarking ``low-order'' methods.
For the sake of completeness, we do provide in the SI the results obtained when selecting the alternative option, i.e., considering all states.

\subsection{Vertical transition energies}

The vertical transition energies obtained with various methods can be found in the SI for all ESs (Table S6).  Obviously, VEEs do not constitute our focus here and we have already commented on similar yet larger sets  previously, \cite{Loo18a} so they have
been included only for the sakes of completeness and reproducibility. A statistical analysis of the results concerning VEEs can be found in Table \ref{Table-1}. For the wave function approaches, the trends are very similar to the previously noted ones with: i) errors
steadily decreasing when increasing the CC excitation order; ii) highly similar behavior for ADC(2) and CC2; and iii) a tendency of ADC(3) to underestimate transition energies. \cite{Loo18a,Loo20b} For TD-DFT, the trends are again similar to these published
elsewhere, \cite{Cas19} all tested XCFs providing underestimated VEEs, the two range-separated hybrids delivering the smallest errors, with SDE and RMSE rather similar to the one obtained with ADC(2) and CC2.  We underline that  the relatively poor 
performance of B3LYP is, at least, partially related to both the small size of the treated molecules and the consideration of many ESs of Rydberg character, the latter being described more faithfully with range-separated hybrids thanks to the introduction of a 
large percentage of exact exchange at large interelectronic distances. \cite{Pea08}

\begin{table}[H]
\caption{Mean Signed Error (MSE), Mean Absolute Error (MAE), Standard Deviation of the Errors (SDE), Root-Mean Square Error (RMSE), Maximal Positive Error [Max($+$)] and Maximal Negative Error [Max($-$)] with respect to the TBEs
for the 46 VEEs listed in Table S6. All values are in eV.}
\label{Table-1}
\begin{footnotesize}
\begin{tabular}{lcccccc}
\hline
\hline
Method			&MSE	&MAE	&SDE	&RMSE	&Max($+$)	&Max($-$)\\
\hline 
CC2				&0.03	&0.21	&0.29	&0.29	&0.60	&-0.71\\
CCSD			&0.10	&0.11	&0.09	&0.14	&0.33	&-0.04\\
CC3				&0.01	&0.03	&0.04	&0.04	&0.13	&-0.05\\
CCSDT			&0.01	&0.02	&0.03	&0.03	&0.09	&-0.06\\
ADC(2)			&0.02	&0.21	&0.29	&0.29	&0.58	&-0.75\\
ADC(3)			&-0.12	&0.20	&0.19	&0.22	&0.40	&-0.40\\
B3LYP			&-0.44	&0.45	&0.47	&0.64	&0.24	&-2.74\\
PBE0			&-0.29	&0.33	&0.40	&0.49	&0.48	&-2.40\\
M06-2X			&-0.27	&0.34	&0.30	&0.41	&0.75	&-0.90\\
CAM-B3LYP		&-0.24	&0.26	&0.22	&0.33	&0.41	&-0.83\\
$\omega$B97X-D	&-0.18	&0.22	&0.22	&0.29	&0.54	&-0.74\\
\hline	
\hline	
\end{tabular}
\end{footnotesize}
\end{table}

\subsection{Ground-state dipoles}

For $\mu^{\mathrm{GS}}$, the most extensive DFT benchmark to date is the one of Hait and Head-Gordon, \cite{Hai18} whereas the investigation of Hodecker \emph{et al.}~already
provides a very valuable assessment of the differences between OU, OR,  and ISR at several levels of theory. \cite{Hod19c} Our goals here are therefore: i) to briefly discuss the quality of the GS dipoles
so that straightforward comparisons with the corresponding ES analysis below can be done; ii) to assess both CC3 and CCSDT against FCI, which has not be done with a reasonable basis set to date; iii)
to provide statistically relevant differences between OU, OR, and ISR. The raw data can be found in Table S7--S9 in the SI.

For the 16 non-centrosymmetric molecules of Figure \ref{Figure-1}, the mean absolute deviations (MADs) obtained when using OU instead of OR are 0.007, 0.004, 0.025, 0.042, and 0.167 D for CCSDT, CC3, CCSD, CC2, and MP2, 
respectively.  For CCSDTQ, a subset of six compounds could be tested (Table S9) and the differences between OU and OR are found to be always insignificant with a MAD of 0.001 D. Consistently with a previous work, \cite{Hod19c}
the ISR-MP2 approach is roughly in-between the OU and OR, with a MAD of 0.090 D as compared to OR. As expected, the impact of neglecting orbital relaxation is sizable with MP2, but rapidly 
decreases when the quality of the wave function is improved to become essentially negligible for all methods incorporating iterative triples. Interestingly, it turns out that CC3 is less sensitive to orbital relaxation than CCSDT, so that 
the trend recently observed for the ground-state dipole of hydrogen fluoride \cite{Hod19c} seems to be valid beyond that specific diatomic molecule.

\begin{figure*}[htp]
\centering
 \includegraphics[width=.8\linewidth]{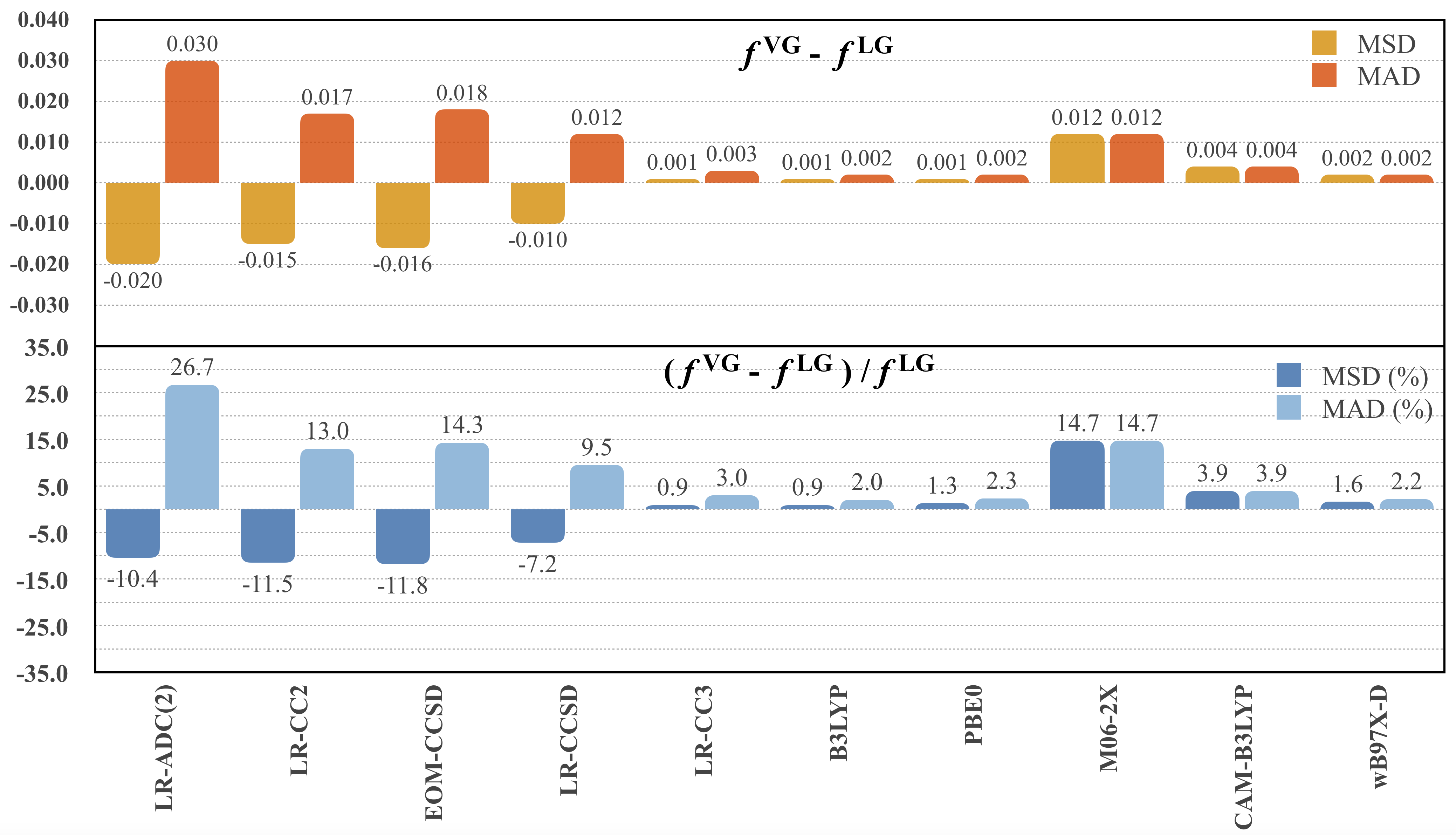}
  \caption{Mean signed deviation (MSD) and mean absolute deviation (MAD) between oscillator strengths in velocity gauge, $f^{\mathrm{VG}}$, and oscillator strengths in length gauge, $f^{\mathrm{LG}}$, for various methods.
  This represents a statistics over 34 oscillator strength values which are listed in the SI.   Top: $f^{\mathrm{VG}} - f^{\mathrm{LG}}$. 
  Bottom: $(f^{\mathrm{VG}} - f^{\mathrm{LG}})/f^{\mathrm{LG}}$ (in percent), where the cases with $f^{\mathrm{LG}} < 0.010$ have been removed in order to provide unbiased statistics.}
   \label{Figure-2}
\end{figure*}

Statistical quantities regarding ground-state dipole's quality (with respect to the present TBEs) for various methods are reported in Table \ref{Table-2}. 

\begin{table}[H]
\caption{Statistical analysis of $\norm*{\mu^{\mathrm{GS}}}$ (in D) for the data listed in Table S7 (16 dipoles). See caption of Table \ref{Table-1} for additional details. }
\label{Table-2}
\begin{footnotesize}
\begin{tabular}{llcccccc}
\hline 
\hline 
Method			&		&MSE	&MAE	&SDE	&RMSE	&Max($+$)	&Max($-$)\\
\hline 
MP2				&OU		&0.15	&0.16	&0.15	&0.21	&0.45	&-0.10	\\
				&ISR		&0.10	&0.14	&0.17	&0.19	&0.50	&-0.22	\\
				&OR		&0.04	&0.08	&0.11	&0.11	&0.23	&-0.16	\\
CC2				&OU		&0.03	&0.12	&0.17	&0.17	&0.31	&-0.33	\\
				&OR		&0.04	&0.09	&0.12	&0.12	&0.23	&-0.22	\\
CCSD			&OU		&0.02	&0.02	&0.02	&0.03	&0.07	&-0.03	\\
				&OR		&0.02	&0.04	&0.04	&0.05	&0.09	&-0.06	\\	
CC3				&OU		&0.00	&0.01	&0.01	&0.01	&0.02	&-0.02	\\
				&OR		&0.00	&0.01	&0.01	&0.01	&0.02	&-0.02	\\
CCSDT			&OU		&0.00	&0.00	&0.01	&0.01	&0.01	&-0.01	\\
				&OR		&0.00	&0.01	&0.01	&0.01	&0.01	&-0.01	\\
B3LYP			&		&0.06	&0.07	&0.06	&0.08	&0.19	&-0.05	\\
PBE0			&		&0.07	&0.07	&0.06	&0.09	&0.21	&-0.03	\\
M06-2X			&		&0.08	&0.09	&0.08	&0.12	&0.29	&-0.09	\\
CAM-B3LYP		&		&0.10	&0.11	&0.07	&0.12	&0.24	&-0.08	\\	
$\omega$B97X-D	&		&0.09	&0.09	&0.07	&0.11	&0.27	&-0.04	\\
\hline 
\hline 
\end{tabular}
\end{footnotesize}
\end{table}

Strikingly, all MSEs are positive (or null) indicating that all tested methods tend to overestimate the FCI  $\norm*{\mu^{\mathrm{GS}}}$  values.  For both MP2 and CC2, 
turning on the orbital relaxation improves the accuracy of the GS dipoles, with typical deviations around 0.1 D only at the OR level. With CCSD, one observes the opposite trend, i.e., OU is on average more accurate than
OR, an effect that we attribute to error compensation.  Finally, for both CC3 and CCSDT, all deviations are negligible, evidencing that both methods estimate $\mu^{\mathrm{GS}}$ with chemical accuracy. Turning now
our attention to DFT, one first notes that all five XCFs yield errors of the same order of magnitude as OR-MP2 and OR-CC2, with a slightly improved consistency (smaller SDE).  For MP2, the same conclusion can be 
easily deduced from the data of Hait and Head-Gordon. \cite{Hai18} These trends strengthen the claim that benchmarking DFT using CC2 (or MP2) reference values is a risky strategy: the latter does not yield smaller errors, at least
for the compact compounds treated herein. Amongst the tested XCFs, B3LYP and PBE0 emerge as the most accurate, but the variations compared to the three other XCFs are  limited and likely not significant

\subsection{Oscillator strengths}

The computed values of the oscillator strength for all tested approaches can be found in Tables S10--S12 in the SI along with the corresponding TBEs, and we discuss below statistical trends only. 

\subsubsection{Impact of the gauge}
An aspect that is hardly discussed in practical applications of ES theories to ``real-life'' compounds is the gauge effect on the magnitude of the oscillator strengths, although comparisons between
theoretical $f$ values and experimental intensities typically guide the identification of the relevant transitions. In Figure \ref{Figure-2}, we provide an overview of the variations of $f$ computed in the
length ($f^{\mathrm{LG}}$) and velocity ($f^{\mathrm{VG}}$) gauges for five wave function methods as well as five XCFs within TD-DFT. The additional figure comparing  length and mixed  
($f^{\mathrm{MG}}$) gauges can be found in Figure S1 in the SI. As expected, the mixed gauge essentially delivers $f$ values bracketed by those obtained with the length and velocity gauges. 
Figure \ref{Figure-2} shows the largest possible differences between the three gauges, but the methodological trends are conserved in other gauge comparisons, e.g., $f^{\mathrm{MG}} - f^{\mathrm{LG}}$.

With the tested wave function approaches, going from $f^{\mathrm{LG}}$ to $f^{\mathrm{VG}}$ induces a decrease of the estimated oscillator strengths in the vast majority of the cases, 
whereas the opposite effect is found within TD-DFT. Following the expected trend, the more accurate the wave function is, the smaller the impact of the gauge is, hence the negligible
gauge variance observed for CC3 with a MAD of 0.003 (i.e., 3\%) between the length and velocity gauges. The gauge effects are more significant for both CCSD and CC2, and even quite large with 
ADC(2) with a MAD one order of magnitude larger than CC3 (0.030 or 27\%).  One also notices from Figure \ref{Figure-2} that the $f$ values determined with the EOM formalism
are more affected by the selected gauge than those computed computed within the LR framework. Turning our attention to TD-DFT, the differences between $f^{\mathrm{LG}}$ and $f^{\mathrm{VG}}$ are generally
small with all XCFs (especially with B3LYP and PBE0) at the notable exception of M06-2X. With the latter functional, variations of ca.~15\%\ are observed between the two gauges, which
is four times larger than with the second most gauge-sensitive XCF (CAM-B3LYP). Nevertheless such effect remains of the same order of magnitude as the one noticed with EOM-CCSD.
As mentioned in the Introduction section of the present paper, the weak gauge sensitivity of TD-DFT might be due to the fulfillment of the Thomas-Reiche-Kuhn sum rule.

As a final note, in Table S18 in the SI, one can find a basis set study of gauge effects for water performed at various levels of theory. Whilst one notes a small decrease of the gauge impact 
when enlarging the basis set, the methodological trends are preserved in going from {\AVTZ} to {\AVFZ}, with significant gauge effects pertaining for both ADC(2) and M06-2X
even with the latter basis set.

\subsubsection{Response \emph{vs} expectation formalisms}

\begin{figure}[htp]
\centering
 \includegraphics[width=.8\linewidth]{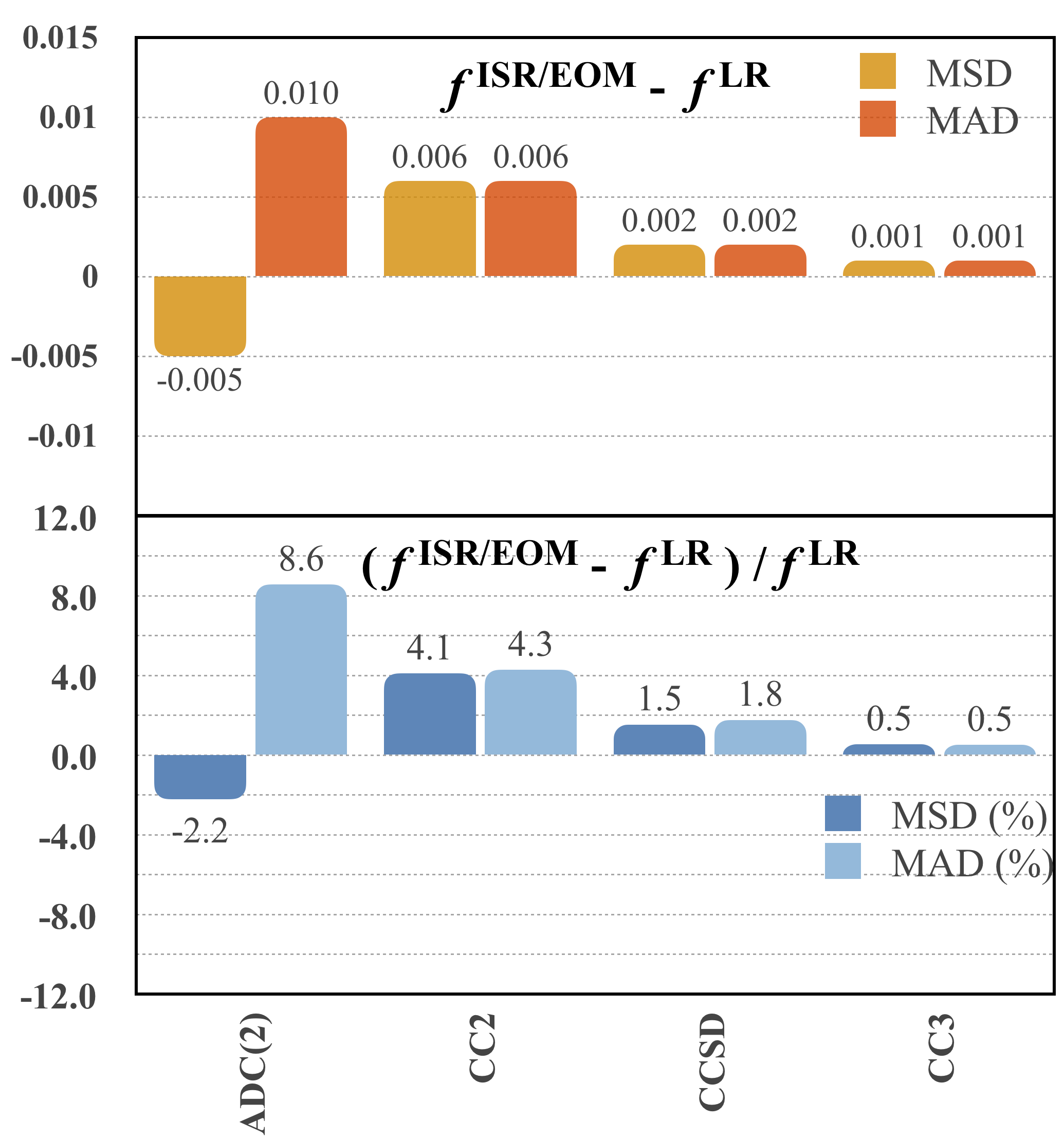}
  \caption{Mean signed deviation (MSD) and mean absolute deviation (MAD) between (length gauge) oscillator strengths computed within the expectation value and linear response formalisms. 
	For ADC(2), we report the difference between ISR and LR, whereas for the three CC methods, the deviations between EOM and LR formalisms are displayed. 
	Top: $f^{\mathrm{ISR/EOM}} - f^{\mathrm{LR}}$. 
	Bottom: $(f^{\mathrm{ISR/EOM}} - f^{\mathrm{LR}})/f^{\mathrm{LR}}$ (in percent), where the cases with $f^{\mathrm{LR}} < 0.010$ have been removed in order to provide unbiased statistics.}
   \label{Figure-3}
\end{figure}

Figure \ref{Figure-3} presents the MSD and MAD for oscillator strengths computed with the ISR and LR formalisms for ADC(2) and the EOM and LR formalisms for CC2, CCSD, and CC3. Again,
one notices that the deviations between the various formalisms are magnified when one considers second-order computational methods. For the CC approaches, EOM generally yields larger oscillator strengths than LR,
but as noted previously, \cite{Car09b,Kan14} these effects are relatively mild with changes of 4\%\ for CC2 and 2\%\ for CCSD, the changes becoming essentially zero with CC3. Of course,
it is likely that the differences between the EOM and LR oscillator strengths would be exacerbated for larger compounds. \cite{Car09b} For ADC(2), the Q-CHEM (ISR) and TURBOMOLE (LR) implementations
yield distinct $f$ estimates, the ISR scheme providing smaller values, with a MAD of 9\%\ between the two schemes. Interestingly, at the ADC(2) level, the impact of the selected formalism 
is about a third of the one of the gauge as one can see by comparing Figures \ref{Figure-2} and \ref{Figure-3}. Test calculations (Table S19) hint that the effects on these formalisms are similar
with the triple- and quadruple-$\zeta$ basis sets.

\subsubsection{Statistical performances}

\begin{figure*}[htp]
\centering
 \includegraphics[width=.96\linewidth]{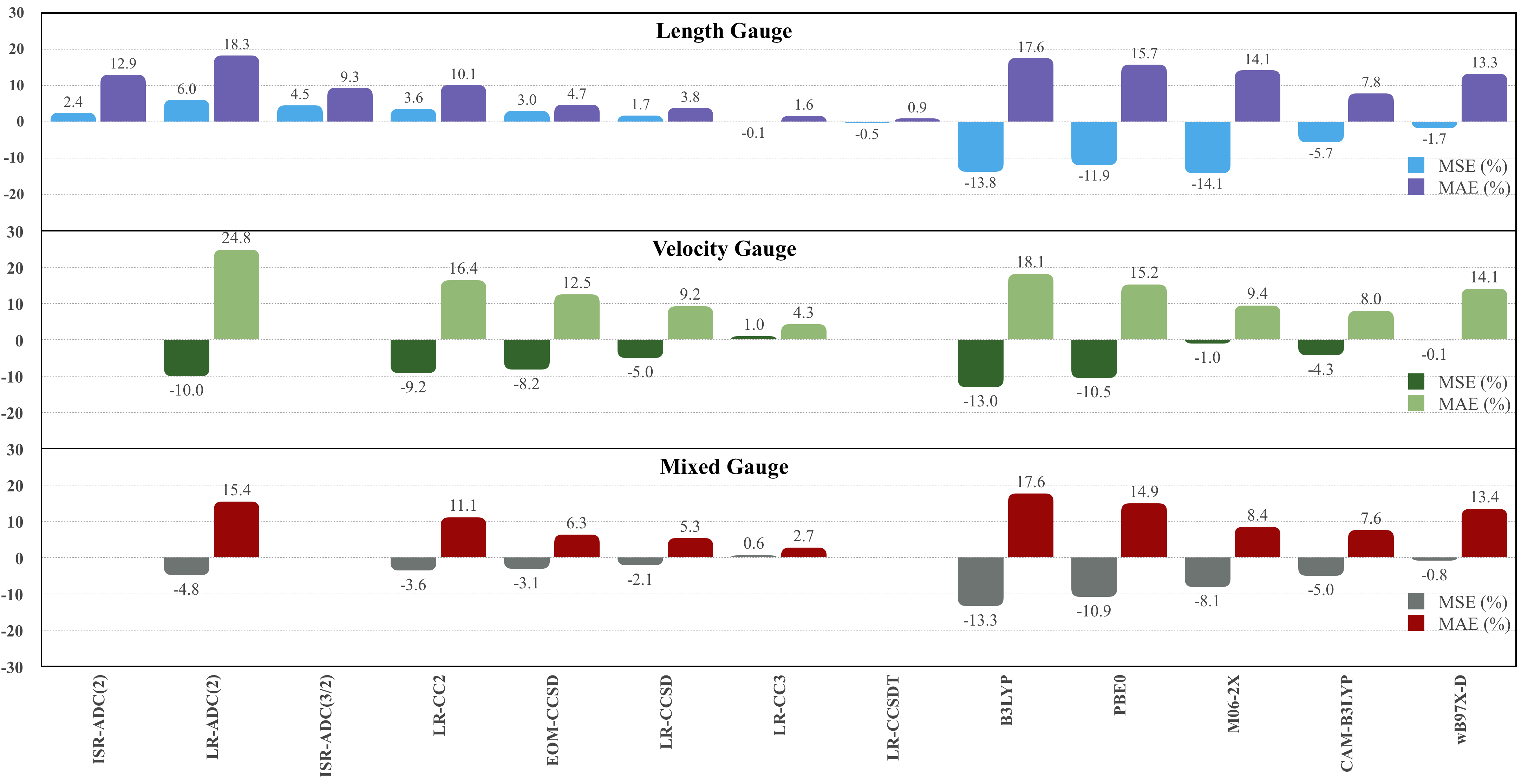}
  \caption{Mean signed error (MSE) and mean absolute error (MAE) (in \%) with respect to the TBEs values for $f^{\mathrm{LG}}$ (top), $f^{\mathrm{VG}}$ (middle), and $f^{\mathrm{MG}}$ (bottom) obtained for various levels of theory. 
  This represents the statistics for 24-out-of-34 $f$ values as we do not include cases with strong state mixing (6 transitions, see Table \ref{Table-3}) nor cases with $f^{\mathrm{TBE}} < 0.010$ (4 transitions). }
   \label{Figure-4}
\end{figure*}

Let us now turn towards comparisons with respect to our TBEs for all tested models, formalisms, and gauges. The complete statistical results can be found in Table \ref{Table-3} whereas Figure \ref{Figure-4} provides
a graphical view of the average relative errors for various methods. The statistical data corresponding to Table \ref{Table-3} are gathered in Table S13 of the SI in which none of the state mixing ES has been dropped out of 
the statistics (see above).

\begin{table}[htp]
\caption{Statistical analysis with respect to TBEs for the oscillator strength $f$ (in absolute values) for the data listed in Table S10--S12 (28-out-of-34 $f$ values, state mixing cases removed). See caption of Table \ref{Table-1} for more details and
Table S13 in the SI for the corresponding analysis obtained for the full set of 34 values.
}
\label{Table-3}
\begin{footnotesize}
\begin{tabular}{lllcccccc}
\hline
\hline
Method			&		&Gauge	&MSE	&MAE	&SDE	&RMSE	&Max($+$)	&Max($-$)\\
\hline 
ADC(2)			&ISR		& LG		&0.004	&0.014	&0.022	&0.026	&0.047	&-0.084	\\
				&LR		& LG		&0.009	&0.020	&0.030	&0.031	&0.098	&-0.089	\\
				&		& VG	&-0.010	&0.023	&0.032	&0.041	&0.043	&-0.118	\\
				&		& MG	&-0.002	&0.014	&0.021	&0.035	&0.044	&-0.085	\\
ADC(3/2)			&ISR		& LG		&0.004	&0.008	&0.013	&0.014	&0.052	&-0.028	\\
CC2				&EOM	& LG		&0.008	&0.014	&0.023	&0.036	&0.084	&-0.078	\\
				& LR		& LG		&0.003	&0.010	&0.018	&0.029	&0.048	&-0.082	\\
				&		& VG	&-0.011	&0.014	&0.020	&0.026	&0.014	&-0.091	\\
				&		& MG	&-0.004	&0.009	&0.016	&0.024	&0.013	&-0.086	\\
CCSD			&EOM	& LG		&0.004	&0.005	&0.008	&0.018	&0.028	&-0.021	\\
				&		&VG		&-0.010	&0.013	&0.019	&0.022	&0.007	&-0.070	\\
				&		& MG	&-0.004	&0.006	&0.009	&0.011	&0.007	&-0.029	\\				
				&LR		& LG		&0.002	&0.004	&0.007	&0.015	&0.019	&-0.023	\\
				&		&VG		&-0.007	&0.009	&0.014	&0.016	&0.008	&-0.051	\\
				&		& MG	&-0.002	&0.005	&0.007	&0.010	&0.008	&-0.028	\\			
CC3				&EOM	& LG		&0.001	&0.002	&0.002	&0.010	&0.011	&-0.002	\\
				&LR		& LG		&0.000	&0.001	&0.002	&0.008	&0.010	&-0.002	\\	
				&		&VG		&0.001	&0.004	&0.005	&0.009	&0.020	&-0.011	\\
				&		& MG	&0.001	&0.002	&0.003	&0.008	&0.011	&-0.005	\\
CCSDT			& LR		& LG		&0.000	&0.001	&0.001	&0.005	&0.003	&-0.005	\\	
B3LYP			&		& LG		&-0.015	&0.017	&0.024	&0.029	&0.007	&-0.110	\\
				&		& VG	&-0.014	&0.017	&0.024	&0.029	&0.015	&-0.108	\\
				&		& MG	&-0.014	&0.017	&0.024	&0.029	&0.009	&-0.109	\\
PBE0			&		& LG		&-0.013	&0.015	&0.021	&0.027	&0.006	&-0.097	\\
				&		& VG	&-0.012	&0.015	&0.021	&0.027	&0.017	&-0.094	\\
				&		& MG	&-0.012	&0.014	&0.021	&0.027	&0.011	&-0.095	\\
M06-2X			&		& LG		&-0.014	&0.014	&0.013	&0.022	&0.000	&-0.047	\\
				&		& VG	&-0.002	&0.009	&0.013	&0.020	&0.045	&-0.029	\\
				&		& MG	&-0.008	&0.008	&0.010	&0.018	&0.002	&-0.034	\\
CAM-B3LYP		&		& LG		&-0.006	&0.007	&0.009	&0.017	&0.004	&-0.032	\\
				&		& VG	&-0.006	&0.008	&0.010	&0.017	&0.013	&-0.032	\\
				&		& MG	&-0.006	&0.007	&0.010	&0.017	&0.007	&-0.032	\\
$\omega$B97X-D	&		& LG		&-0.006	&0.011	&0.016	&0.021	&0.015	&-0.056	\\
				&		& VG	&-0.005	&0.011	&0.016	&0.022	&0.027	&-0.050	\\
				&		& MG	&-0.006	&0.011	&0.016	&0.022	&0.021	&-0.053	\\				
\hline	
\hline	
\end{tabular}
\end{footnotesize}
\end{table}

A first general conclusion is that all methods do deliver rather consistent estimates (small SDEs) with reasonably small absolute and relative errors. Indeed, the largest MAE reported in Figure \ref{Figure-4} is smaller
than 25\%. In the commonly-selected length gauge, the MSE obtained with all tested wave function approaches are positive, whereas all investigated TD-DFT models tend to undershoot $f^{\mathrm{TBE}}$. Remaining 
in the length gauge, the ISR representation of ADC(2) outperforms the LR formulation, and ADC(3/2) provides a significant improvement over ADC(2), which contrasts with the vertical transition energies. We therefore confirm 
the earlier conclusion that ADC(3/2) is superior to ADC(2) for $f$. \cite{Har14} In the CC family, one notes that both CCSDT and CC3 provide extremely small deviations with respect to the TBEs, and, consistently with the analysis made in the 
previous paragraph, selecting EOM or LR has no significant effect for CC3. Concerning CCSD, the LR representation has a slight edge, an effect that is somewhat enlarged in CC2. These conclusions are consistent
with an earlier work that reported CC2 errors significantly larger than their CCSD counterparts, although remaining \emph{``qualitatively correct''}. \cite{Kan14}  Amongst the five tested XCFs, CAM-B3LYP is clearly the 
best performer in the length gauge with a MAE of 8\%\ only, whereas all other XCFs yield MAEs in the 13--17\%\ range (see  Figure \ref{Figure-4}).  Again, it is clear that CAM-B3YP outperforms both CC2 and ADC(2) for each statistical quantity listed
in Table \ref{Table-3}.  In the same vein, the PBE0 errors are quite comparable to those obtained with EOM-CC2. As we noted above for $\mu^\mathrm{GS}$, one should then be cautious when using a second-order
approach to evaluate the performance of TD-DFT.  Let us now turn towards the velocity gauge. The accuracy of the wave function results seem to systematically degrade when turning to this gauge with negative MSEs and larger MAEs
than with the length gauge, whereas the TD-DFT statistics are mostly unaffected, except for a significant improvement of M06-2X. The mixed gauge results are more contrasted, with an improvement for ADC(2) and M06-2X, no major changes for the other
XCFs (as compared to $f^{\mathrm{LG}}$), and slight worsening of the errors associated with the CC methods.

\subsection{Excited-state dipoles}

Excited-state dipole moments computed with various methods are listed in Tables S14 and S15 in the SI. At this stage, we recall that we have considered in our statistical signed analyses the dipole norm,
so that the reported MSE are indicative of amplitudes, not directions. Moreover, we have removed six ESs strongly influenced by state mixing Subsec.~\ref{sec:stat}.

\subsubsection{On the impact of orbital relaxation in LR}

\begin{figure}[htp]
\centering
 \includegraphics[width=1.0\linewidth]{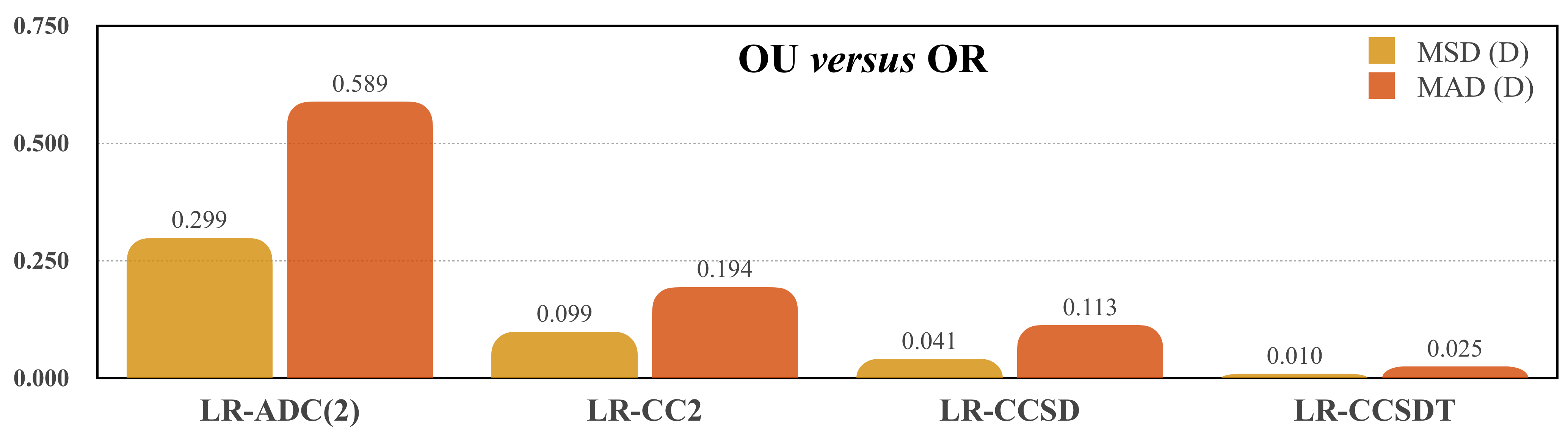}
  \caption{Mean signed deviation (MSD) and mean absolute deviation (MAD) between the OU and OR ES dipoles ($\mu^\mathrm{OU}-\mu^\mathrm{OR}$, in D) computed with various methods. 
  This represents the statistics for 40 ${\mu^{\mathrm{ES}}}$ values listed in the SI. }
   \label{Figure-5}
\end{figure}

Let us first discuss the impact of various relaxation schemes on the computed ${\mu^{\mathrm{ES}}}$. First, from Table S21 in the SI, it can be concluded that the orbital relaxation effects depend (much)
more on the considered excited state and method than on the selected atomic basis set. A graphical analysis is given in Figure \ref{Figure-5}. When neglecting orbital
relaxation in CC theory, one notices a slight overestimation of $\norm*{\mu^{\mathrm{ES}}}$, though both negative and positive deviations are observed when investigating individual cases. Of course, and as expected, 
the difference between OR and OU decreases when increasing the CC order: it is significant for CC2 (0.19 D), smaller for CCSD (0.11 D), and negligible for CCSDT 
(0.03 D). For our set, using LR(OU)-ADC(2) yields a MAD of 0.59 D as compared to LR(OR)-ADC(2). One noteworthy point is that the statistical OR/OU discrepancies are 
much larger for $\mu^{\mathrm{ES}}$ than for $\mu^{\mathrm{GS}}$, irrespective of the selected wave function model. 
Indeed, for the GS dipoles, we found MADs of 0.01 D for CCSDT, 0.03 D for CCSD, 0.04 D for CC2 and 0.17 D for MP2 (\emph{vide supra}) between OU and OR, all values being roughly 
one-third of their ES analogs displayed in Figure \ref{Figure-5}. In other words, the ES dipoles are approximatively three times more sensitive to orbital relaxation than the GS dipoles.

\subsubsection{Response \emph{vs} expectation formalisms}

\begin{figure}[htp]
\centering
 \includegraphics[width=1.0\linewidth]{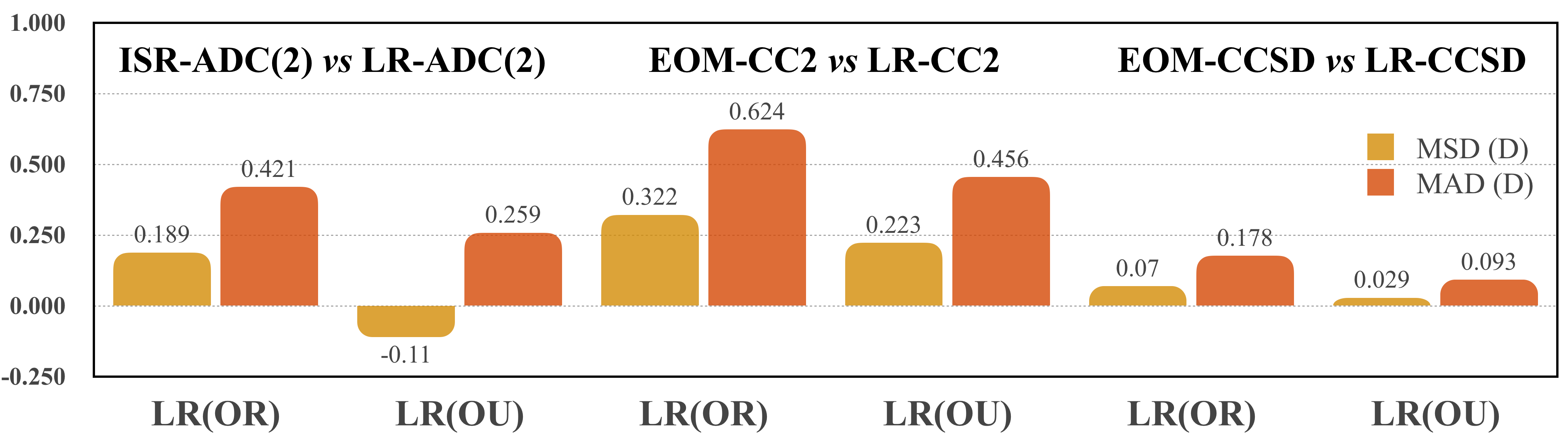}
  \caption{Mean signed deviation (MSD) and mean absolute deviation (MAD) between ES dipoles computed within the expectation value formalism for ADC(2), CC2, and CCSD and their LR(OR) or LR(OU) counterparts.   
  This represents the statistics of 40 ${\mu^{\mathrm{ES}}}$ values listed in the SI. }
   \label{Figure-6}
\end{figure}

Let us now compare the ES dipoles obtained with the EOM and LR formalisms for both CC2 and CCSD. The raw data can be found in Table S14 in the SI, whereas a statistical analysis can
be found in Figure \ref{Figure-6}. We found that, typically, the EOM-CCSD dipoles are more similar to the LR(OU)-CCSD values (MAD of 0.09 D) than the
LR(OR)-CCSD data (MAD of 0.18 D), the magnitude of the ES dipoles,  $\norm*{\mu^{\mathrm{ES}}}$, tending to be slightly larger with the EOM formalism.  If one turns to 
CC2, one observes exactly the same trends, but the deviations between EOM-CC2 and LR-CC2 are much larger with MAD of 0.46 D (OU) and 0.62 D (OR). Turning our attention to 
ADC(2), we notice that ISR,  while providing dipole moments in between OU and OR, yields ES dipole magnitude typically closer to the former. The underlying reasons for this last outcome have 
been unveiled elsewhere. \cite{Hod19c}  The absolute deviation between ISR-ADC(2) $\mu^{\mathrm{ES}}$ and their LR counterparts averages to 0.26 D (OU) and 0.42 D (OR).

\subsubsection{Statistical performances}
\label{sec:stat}
\begin{figure*}[htp]
\centering
 \includegraphics[width=\linewidth]{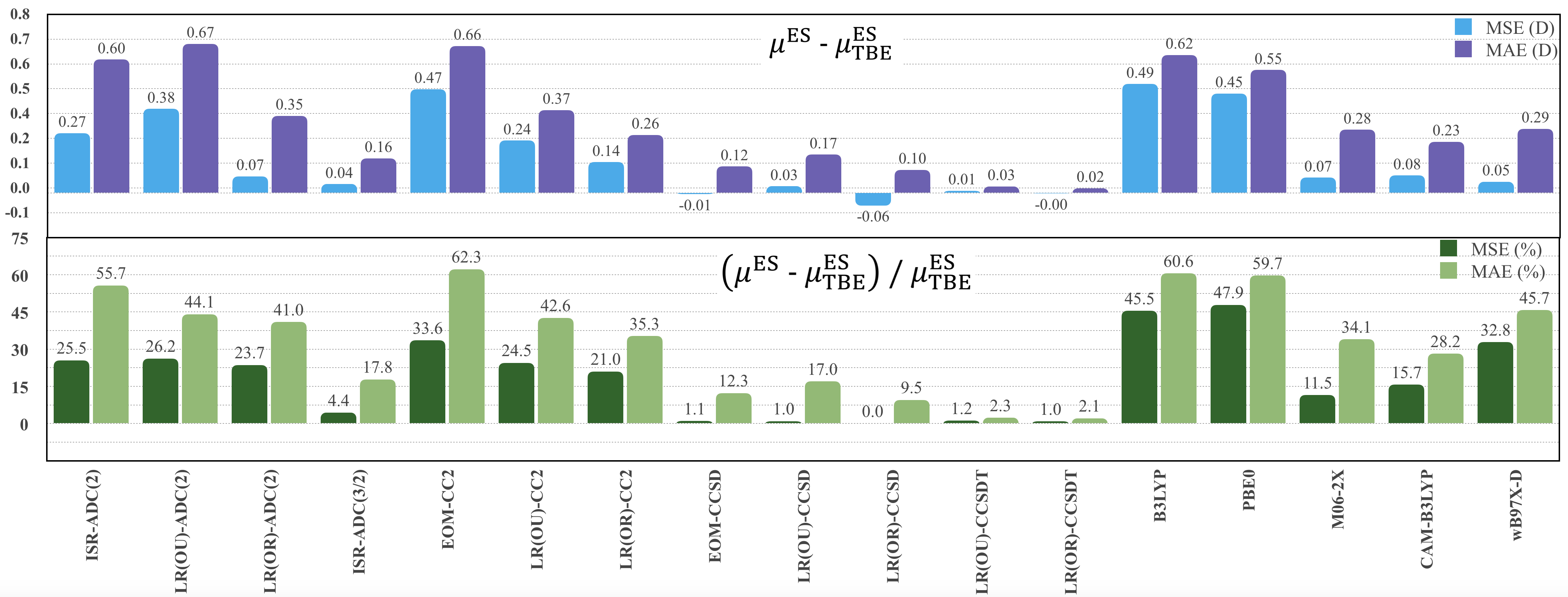}
  \caption{Mean signed error (MSE) and mean absolute error (MAE) with respect to the $\mu^{\mathrm{ES}}$ TBE values obtained for various levels of theory. This represents the statistics of 37-out-of-40 dipoles. See caption of Table \ref{Table-4} 
  for more details.  Top: $\mu^{\mathrm{ES}} - \mu_\text{TBE}^{\mathrm{ES}}$ (in D).  Bottom: $(\mu^{\mathrm{ES}} - \mu_\text{TBE}^{\mathrm{ES}})/\mu_\text{TBE}^{\mathrm{ES}}$ (in percent), where the cases with $\norm*{\mu_\text{TBE}^{\mathrm{ES}}} < 0.1$ D 
  have been removed in order to provide unbiased statistics.}
   \label{Figure-7}
\end{figure*}

Finally, let us compare to our TBEs  the ES dipole moments determined with the twelve wave function methods and five XCFs. The results are listed in Table \ref{Table-4}
and a selection of values are displayed in Figure \ref{Figure-7}. Before discussing these results, it is important to recall that all our calculations are performed with the \emph{aug}-cc-pVTZ basis set. As can be seen in Section
S6.5 in the SI for the test case of water, the impact of increasing the basis set size on the estimated ES dipole moments is rather uniform within the wave function approaches, but might significantly differ
between wave function and TD-DFT approaches. In other words, the TD-DFT errors reported below could be significantly different if another reference basis set was selected.
From Table  \ref{Table-4}, one notices, except for some CCSD variants, that all tested approaches have positive MSEs, i.e., they tend to overestimate the 
magnitude of the dipole moments, which parallels the finding obtained for $\norm*{\mu^{\mathrm{GS}}}$.  When improving the level of theory by increasing the expansion order, one clearly improves the 
accuracy of the $\mu^{\mathrm{ES}}$ values obtained via the wave function methods, with MAEs of 0.60 and 0.16 D with ADC(2) and ADC(3/2) in the ISR formalism, and MAEs of 0.26, 0.11, and
0.02 D for LR-CC2, LR-CCSD, and LR-CCSDT, respectively, when orbital relaxation is accounted for. It is therefore clear, at least for the present set of small molecules, that while ADC(3) transition 
energies are of similar accuracy as their ADC(2) counterparts, going up one rung on the ADC ladder yields a very significant improvement for properties (including ES dipoles), ADC(3/2) clearly 
outperforming CC2 and providing an accuracy comparable to the one obtained with CCSD.  For this latter method, and consistently with the above, there is a rather small difference in accuracy 
between the EOM and the two LR variants, with a MAD of 0.12 D for EOM-CCSD, and 0.10 D for LR(OR)-CCSD. In contrast for both LR-ADC(2) and LR-CC2, accounting for orbital relaxation does 
not only change significantly the values as stated above, but also markedly improves the accuracy, with errors halved as compared to LR(OU)-ADC(2) for the ADC approach(Figure \ref{Figure-7}).
The ISR-ADC(2) $\mu^{\mathrm{ES}}$ show an accuracy slightly better than LR(OU)-ADC(2). By comparing the data listed in Tables \ref{Table-2} and \ref{Table-4}, it also appears that the typical 
absolute errors are roughly tripled going from GS to ES properties, irrespective of the wave function method.  We note that our MAE for both LR(OR)-ADC(2) and LR(OR)-CC2 are roughly 50\%\ larger 
than the ones reported by Hellweg \cite{Hel11} from comparisons with experiments. This is likely related to the consideration of higher-lying Rydberg ESs here for which no experimental values are 
available. Indeed, it is known that both ADC(2) and CC2 are significantly more robust for valence than Rydberg transitions.\cite{Kan17,Loo18a}

\begin{table}[htp]
\caption{Statistical analysis for $\mu^{\mathrm{ES}}$ (in D) for the data listed in Table S14 and S15 (37-out-of-40 dipoles, state mixing cases removed). See caption of Table \ref{Table-1} for more details and 
Table S17 in the SI for the corresponding analysis obtained for the full set. Note that the MSE, Max($+$), and Max($-$) values are obtained by considering the norm of the dipole moments, $\norm*{\mu^{\mathrm{ES}}}$ 
whereas MAEs, SDEs, and RMSEs take into account the sign of the dipole.}
\label{Table-4}
\begin{footnotesize}
\begin{tabular}{llcccccc}
\hline
\hline
Method			&		&MSE	&MAE	&SDE	&RMSE	&Max($+$)&Max($-$)\\
\hline 
ADC(2)			&	ISR	&0.268	&0.598	&0.749	&0.795	&2.195	&-1.778	\\
				& 	OU	&0.377	&0.668	&0.812	&0.903	&2.384	&-2.041	\\
				&	OR	&0.074	&0.345	&0.519	&0.536	&1.384	&-1.866	\\
ADC(3/2)			&	ISR	&0.040	&0.155	&0.229	&0.237	&0.360	&-0.523	\\
CC2				&	EOM	&0.465	&0.658	&0.743	&0.865	&2.395	&-0.874	\\
				&	OU	&0.235	&0.371	&0.508	&0.525	&1.915	&-0.819	\\
				&	OR	&0.138	&0.260	&0.375	&0.388	&1.419	&-0.692	\\
CCSD			&	EOM	&-0.005	&0.119	&0.176	&0.186	&0.221	&-0.633	\\
				&	OU	&0.030	&0.172	&0.217	&0.246	&0.604	&-0.748	\\
				&	OR	&-0.056	&0.103	&0.175	&0.180	&0.122	&-0.638	\\
CCSDT			&	OU	&0.010	&0.029	&0.044	&0.047	&0.153	&-0.130	\\
				&	OR	&-0.002	&0.021	&0.037	&0.038	&0.095	&-0.135	\\
B3LYP			&		&0.488	&0.618	&1.031	&1.059	&3.066	&-0.570	\\
PBE0			&		&0.445	&0.551	&0.899	&0.924	&2.570	&-0.603	\\
M06-2X			&		&0.069	&0.283	&0.392	&0.402	&1.479	&-0.655	\\
CAM-B3LYP		&		&0.078	&0.229	&0.292	&0.308	&0.786	&-0.463	\\
$\omega$B97X-D	&		&0.050	&0.287	&0.421	&0.450	&0.965	&-1.249	\\
\hline	
\hline	
\end{tabular}
\end{footnotesize}
\end{table}

If we now turn our attention to the five tested TD-DFT approaches,  two sets of XCFs clearly emerge. On the one hand, B3LYP and PBE0 very significantly overshoot the magnitude of the
ES dipoles, with large MAEs of 0.62 and 0.55 D, respectively. While these errors are quite comparable to the ones obtained with ADC(2) in its LR(OU) or ISR formulations, they are almost one order
of magnitude larger than those for the GS dipoles (0.07 D for both hybrids). As our set is constituted of small molecules, for which one does not expect charge-transfer effects to be important, this is
likely related to the consideration of high-lying ESs.  Indeed, as already stated above, global hybrids with low exact exchange are known to be less adequate for Rydberg excitations due to the 
wrong behavior of the exchange-correlation kernel at large interelectronic distances. \cite{Cas98,Toz98} Nevertheless, for B3LYP, Thiel's group reported a MAE of 0.59 D as compared to CASPT2 on the basis of 
numerous valence ESs, \cite{Sch08} a value very similar to our MAE.   On the other hand, one finds M06-2X, CAM-B3LYP and $\omega$B97X-D with significantly smaller MAEs of ca.~0.25 D, 
that is roughly three times the corresponding deviations associated with $\mu^{\mathrm{GS}}$, and about the same order of magnitude as the one obtained with LR(OR)-CC2. Amongst the five tested XCFs, 
CAM-B3LYP seems to have the edge in terms of both absolute accuracy and consistency. Following the GS analysis of Ref.~\citenum{Hai18}, one can thus conclude that CAM-B3LYP does provide 
a more faithful description of the ES density than the other four functionals.

\section{Conclusions and outlook}

In this benchmark study, we have considered a set of 46 vertical transition energies, 16 GS dipoles, 34 oscillator strengths, and 40 ES dipole moments of very high quality to assess a series of single-reference
wave function methods and five XCFs within TD-DFT. The TBEs used as references are obtained with the \emph{aug}-cc-pVTZ basis set and all contain at least CCSDTQ or CCSDTQP corrections, which makes them of near-FCI quality.
For the transition energies, our conclusions are in line with previous works with typical errors of 0.2 eV for ADC(2), ADC(3), CC2, but negligible deviations with CC3 and CCSDT, and 
errors in the 0.22--0.45 eV range for TD-DFT depending on the functional, B3LYP being the least accurate of the five we considered for the present set. For the GS dipoles, we found, as expected, that accounting for orbital
relaxation is significant at the MP2 level (changes of ca.~0.17 D) but rapidly becomes irrelevant as the CC excitation order increases with deviations as small as 0.01 D with CC3 and CCSDT. All five considered functionals 
yield errors around 0.10 D only, and appear to be of similar accuracy as MP2 for GS dipoles.  For the oscillator strengths, the mixed gauge typically yields results in between the velocity and length gauges.
The differences between the two latter can be rather large with, e.g., MADs of 27\%\ with ADC(2), 15\%\ for M06-2X, and 13\%\ for LR-CC2. In comparison, the changes induced by going from
EOM to LR at the CC2 level are smaller (4\%\ only). In terms of accuracy, the MAEs steadily decrease when improving the expansion order, e.g., the ADC(3/2) $f$ are more accurate than their ADC(2) counterparts. CAM-B3LYP
appears to be the most accurate XCFs for the oscillator strengths with a MAE of 8\%\ or 0.008, noticeably outperforming LR-CC2. For the ES dipole moments, the influence of
orbital relaxation as well as the typical error bars given by the  wave function approaches are approximatively three times larger than for the GS. Allowing orbital relaxation improves significantly the LR-ADC(2) and
LR-CC2 estimates. Notably, ISR ADC(3/2) ES dipoles are much more accurate than their ISR ADC(2) counterparts. In TD-DFT, the results are contrasted: B3LYP and PBE0 overestimate strongly
the (magnitude of the) ES dipoles, whereas the three other functionals (CAM-B3LYP in particular) yield acceptable deviations in the range 0.2--0.3 D, comparable to LR(OR)-CC2's deviations, and clearly smaller than EOM-CC2's.

Finally, and in line with our first mountaineering paper, \cite{Loo18a} while the use of chemically-accurate reference values of near-FCI quality provides a definitive answer regarding the accuracy of 
various approaches for a given basis set (here: \emph{aug}-cc-pVTZ), one is, of course, left wondering if the present outcomes would pertain for larger, more extended compounds. While there is no crystal-clear answer to this question at this stage,
the present effort demonstrates that CC theory including triples (CC3 and CCSDT) provide oscillator strengths and dipoles that can be viewed as near-flawless, e.g., the EOM-CC3 scheme provides $f$ values with
a MAE of 0.002 only whereas LR-CCSDT delivers $\mu^{\mathrm{ES}}$ values typically within 0.05 D of the corresponding TBEs. Therefore, for medium-sized molecules at least, these models can likely be used as 
trustworthy references. Within this framework, we recall that extensive sets of (LR-)CC3 oscillator strengths are already available for medium-sized molecules.\cite{Kan17,Loo18a,Loo20a,Loo20d}

\section*{Acknowledgements}
PFL and DJ thank Michael Herbst and Trond Saue for fruitful discussions. RS and DJ are indebted to the R\'egion des Pays de la Loire program for support in the framework of the Opt-Basis grant. 
PFL thanks the European Research Council (ERC) under the European Union's Horizon 2020 research and innovation programme (Grant agreement No. 863481) for financial support.
This work benefited from the support of the ANR in the framework of the PIA program ANR-18-EURE-0012.
This research used computational resources of (i) the GENCI-TGCC (Grant No.~2019-A0060801738);  (ii) CALMIP under allocation 
2020-18005 (Toulouse); (iii) CCIPL (\emph{Centre de Calcul Intensif des Pays de Loire});  (iv) a local Troy cluster and (v) HPC resources from ArronaxPlus  (grant ANR-11-EQPX-0004 funded by the 
French National Agency for Research). 

\section*{Supporting Information Available}
Additional reference data for five compounds. Raw values for transition energies, oscillator strengths, and dipole moments. Basis set effects for water. Cartesian coordinates. 
The Supporting Information is available free of charge at https://pubs.acs.org/doi/10.1021/{doi}.

\bibliography{biblio-new}

\providecommand{\latin}[1]{#1}
\makeatletter
\providecommand{\doi}
  {\begingroup\let\do\@makeother\dospecials
  \catcode`\{=1 \catcode`\}=2 \doi@aux}
\providecommand{\doi@aux}[1]{\endgroup\texttt{#1}}
\makeatother
\providecommand*\mcitethebibliography{\thebibliography}
\csname @ifundefined\endcsname{endmcitethebibliography}
  {\let\endmcitethebibliography\endthebibliography}{}
\begin{mcitethebibliography}{135}
\providecommand*\natexlab[1]{#1}
\providecommand*\mciteSetBstSublistMode[1]{}
\providecommand*\mciteSetBstMaxWidthForm[2]{}
\providecommand*\mciteBstWouldAddEndPuncttrue
  {\def\EndOfBibitem{\unskip.}}
\providecommand*\mciteBstWouldAddEndPunctfalse
  {\let\EndOfBibitem\relax}
\providecommand*\mciteSetBstMidEndSepPunct[3]{}
\providecommand*\mciteSetBstSublistLabelBeginEnd[3]{}
\providecommand*\EndOfBibitem{}
\mciteSetBstSublistMode{f}
\mciteSetBstMaxWidthForm{subitem}{(\alph{mcitesubitemcount})}
\mciteSetBstSublistLabelBeginEnd
  {\mcitemaxwidthsubitemform\space}
  {\relax}
  {\relax}

\bibitem[Gonz{\'a}lez \latin{et~al.}(2012)Gonz{\'a}lez, Escudero, and
  Serrano-Andr\`es]{Gon12}
Gonz{\'a}lez,~L.; Escudero,~D.; Serrano-Andr\`es,~L. Progress and Challenges in
  the Calculation of Electronic Excited States. \emph{ChemPhysChem}
  \textbf{2012}, \emph{13}, 28--51\relax
\mciteBstWouldAddEndPuncttrue
\mciteSetBstMidEndSepPunct{\mcitedefaultmidpunct}
{\mcitedefaultendpunct}{\mcitedefaultseppunct}\relax
\EndOfBibitem
\bibitem[Dreuw and Head-Gordon(2005)Dreuw, and Head-Gordon]{Dre05}
Dreuw,~A.; Head-Gordon,~M. Single-Reference \emph{ab initio} Methods for the
  Calculation of Excited States of Large Molecules. \emph{Chem. Rev.}
  \textbf{2005}, \emph{105}, 4009--4037\relax
\mciteBstWouldAddEndPuncttrue
\mciteSetBstMidEndSepPunct{\mcitedefaultmidpunct}
{\mcitedefaultendpunct}{\mcitedefaultseppunct}\relax
\EndOfBibitem
\bibitem[Adamo and Jacquemin(2013)Adamo, and Jacquemin]{Ada13a}
Adamo,~C.; Jacquemin,~D. The calculations of Excited-State Properties with
  Time-Dependent Density Functional Theory. \emph{Chem. Soc. Rev.}
  \textbf{2013}, \emph{42}, 845--856\relax
\mciteBstWouldAddEndPuncttrue
\mciteSetBstMidEndSepPunct{\mcitedefaultmidpunct}
{\mcitedefaultendpunct}{\mcitedefaultseppunct}\relax
\EndOfBibitem
\bibitem[Laurent and Jacquemin(2013)Laurent, and Jacquemin]{Lau13}
Laurent,~A.~D.; Jacquemin,~D. TD-DFT Benchmarks: A Review. \emph{Int. J.
  Quantum Chem.} \textbf{2013}, \emph{113}, 2019--2039\relax
\mciteBstWouldAddEndPuncttrue
\mciteSetBstMidEndSepPunct{\mcitedefaultmidpunct}
{\mcitedefaultendpunct}{\mcitedefaultseppunct}\relax
\EndOfBibitem
\bibitem[Roos \latin{et~al.}(2007)Roos, Andersson, F{\"u}lscher, Malmqvist,
  Serrano-Andres, Pierloot, and Merchan]{Roo07}
Roos,~B.~O.; Andersson,~K.; F{\"u}lscher,~M.~P.; Malmqvist,~P.-a.;
  Serrano-Andres,~L.; Pierloot,~K.; Merchan,~M. \emph{Advances in Chemical
  Physics}; John Wiley {\&} Sons, Inc., 2007; Vol.~93; Chapter 5, pp
  219--331\relax
\mciteBstWouldAddEndPuncttrue
\mciteSetBstMidEndSepPunct{\mcitedefaultmidpunct}
{\mcitedefaultendpunct}{\mcitedefaultseppunct}\relax
\EndOfBibitem
\bibitem[Krylov(2006)]{Kry06}
Krylov,~A.~I. Spin-Flip Equation-of-Motion Coupled-Cluster Electronic Structure
  Method for a Description of Excited States, Bond Breaking, Diradicals, and
  Triradicals. \emph{Acc. Chem. Res.} \textbf{2006}, \emph{39}, 83--91\relax
\mciteBstWouldAddEndPuncttrue
\mciteSetBstMidEndSepPunct{\mcitedefaultmidpunct}
{\mcitedefaultendpunct}{\mcitedefaultseppunct}\relax
\EndOfBibitem
\bibitem[Piecuch \latin{et~al.}(2002)Piecuch, Kowalski, Pimienta, and
  Mcguire]{Pie02}
Piecuch,~P.; Kowalski,~K.; Pimienta,~I. S.~O.; Mcguire,~M.~J. Recent Advances
  in Electronic Structure Theory: Method of Moments of Coupled-Cluster
  Equations and Renormalized Coupled-Cluster Approaches. \emph{Int. Rev. Phys.
  Chem.} \textbf{2002}, \emph{21}, 527--655\relax
\mciteBstWouldAddEndPuncttrue
\mciteSetBstMidEndSepPunct{\mcitedefaultmidpunct}
{\mcitedefaultendpunct}{\mcitedefaultseppunct}\relax
\EndOfBibitem
\bibitem[Sneskov and Christiansen(2012)Sneskov, and Christiansen]{Sne12}
Sneskov,~K.; Christiansen,~O. Excited State Coupled Cluster Methods.
  \emph{WIREs Comput. Mol. Sci.} \textbf{2012}, \emph{2}, 566--584\relax
\mciteBstWouldAddEndPuncttrue
\mciteSetBstMidEndSepPunct{\mcitedefaultmidpunct}
{\mcitedefaultendpunct}{\mcitedefaultseppunct}\relax
\EndOfBibitem
\bibitem[Ghosh \latin{et~al.}(2018)Ghosh, Verma, Cramer, Gagliardi, and
  Truhlar]{Gho18}
Ghosh,~S.; Verma,~P.; Cramer,~C.~J.; Gagliardi,~L.; Truhlar,~D.~G. Combining
  Wave Function Methods with Density Functional Theory for Excited States.
  \emph{Chem. Rev.} \textbf{2018}, \emph{118}, 7249--7292\relax
\mciteBstWouldAddEndPuncttrue
\mciteSetBstMidEndSepPunct{\mcitedefaultmidpunct}
{\mcitedefaultendpunct}{\mcitedefaultseppunct}\relax
\EndOfBibitem
\bibitem[Blase \latin{et~al.}(2020)Blase, Duchemin, Jacquemin, and Loos]{Bla20}
Blase,~X.; Duchemin,~I.; Jacquemin,~D.; Loos,~P.~F. The Bethe-Salpeter
  Formalism: From Physics to Chemistry. \emph{J. Phys. Chem. Lett.}
  \textbf{2020}, \emph{11}, 7371\relax
\mciteBstWouldAddEndPuncttrue
\mciteSetBstMidEndSepPunct{\mcitedefaultmidpunct}
{\mcitedefaultendpunct}{\mcitedefaultseppunct}\relax
\EndOfBibitem
\bibitem[Loos \latin{et~al.}(2020)Loos, Scemama, and Jacquemin]{Loo20c}
Loos,~P.~F.; Scemama,~A.; Jacquemin,~D. The Quest for Highly-Accurate
  Excitation Energies: A Computational Perspective. \emph{J. Phys. Chem. Lett.}
  \textbf{2020}, \emph{11}, 2374--2383\relax
\mciteBstWouldAddEndPuncttrue
\mciteSetBstMidEndSepPunct{\mcitedefaultmidpunct}
{\mcitedefaultendpunct}{\mcitedefaultseppunct}\relax
\EndOfBibitem
\bibitem[Runge and Gross(1984)Runge, and Gross]{Run84}
Runge,~E.; Gross,~E. K.~U. Density-Functional Theory for Time-Dependent
  Systems. \emph{Phys. Rev. Lett.} \textbf{1984}, \emph{52}, 997--1000\relax
\mciteBstWouldAddEndPuncttrue
\mciteSetBstMidEndSepPunct{\mcitedefaultmidpunct}
{\mcitedefaultendpunct}{\mcitedefaultseppunct}\relax
\EndOfBibitem
\bibitem[Casida(1995)]{Cas95}
Casida,~M.~E. In \emph{Time-Dependent Density-Functional Response Theory for
  Molecules}; Chong,~D.~P., Ed.; Recent Advances in Density Functional Methods;
  World Scientific: Singapore, 1995; Vol.~1; pp 155--192\relax
\mciteBstWouldAddEndPuncttrue
\mciteSetBstMidEndSepPunct{\mcitedefaultmidpunct}
{\mcitedefaultendpunct}{\mcitedefaultseppunct}\relax
\EndOfBibitem
\bibitem[Schirmer(1982)]{Sch82}
Schirmer,~J. Beyond the Random-Phase Approximation: a new Approximation Scheme
  for the Polarization Propagator. \emph{Phys. Rev. A} \textbf{1982},
  \emph{26}, 2395--2416\relax
\mciteBstWouldAddEndPuncttrue
\mciteSetBstMidEndSepPunct{\mcitedefaultmidpunct}
{\mcitedefaultendpunct}{\mcitedefaultseppunct}\relax
\EndOfBibitem
\bibitem[Trofimov \latin{et~al.}(1999)Trofimov, Stelter, and Schirmer]{Tro99}
Trofimov,~A.~B.; Stelter,~G.; Schirmer,~J. A Consistent Third-Order Propagator
  Method for Electronic Excitation. \emph{J. Chem. Phys.} \textbf{1999},
  \emph{111}, 9982--9999\relax
\mciteBstWouldAddEndPuncttrue
\mciteSetBstMidEndSepPunct{\mcitedefaultmidpunct}
{\mcitedefaultendpunct}{\mcitedefaultseppunct}\relax
\EndOfBibitem
\bibitem[Harbach \latin{et~al.}(2014)Harbach, Wormit, and Dreuw]{Har14}
Harbach,~P. H.~P.; Wormit,~M.; Dreuw,~A. The Third-Order Algebraic Diagrammatic
  Construction Method (ADC(3)) for the Polarization Propagator for Closed-Shell
  Molecules: Efficient Implementation and Benchmarking. \emph{J. Chem. Phys.}
  \textbf{2014}, \emph{141}, 064113\relax
\mciteBstWouldAddEndPuncttrue
\mciteSetBstMidEndSepPunct{\mcitedefaultmidpunct}
{\mcitedefaultendpunct}{\mcitedefaultseppunct}\relax
\EndOfBibitem
\bibitem[Dreuw and Wormit(2015)Dreuw, and Wormit]{Dre15}
Dreuw,~A.; Wormit,~M. The Algebraic Diagrammatic Construction Scheme for the
  Polarization Propagator for the Calculation of Excited States. \emph{WIREs
  Comput. Mol. Sci.} \textbf{2015}, \emph{5}, 82--95\relax
\mciteBstWouldAddEndPuncttrue
\mciteSetBstMidEndSepPunct{\mcitedefaultmidpunct}
{\mcitedefaultendpunct}{\mcitedefaultseppunct}\relax
\EndOfBibitem
\bibitem[Christiansen \latin{et~al.}(1995)Christiansen, Koch, and
  J{\o}rgensen]{Chr95}
Christiansen,~O.; Koch,~H.; J{\o}rgensen,~P. The Second-Order Approximate
  Coupled Cluster Singles and Doubles Model CC2. \emph{Chem. Phys. Lett.}
  \textbf{1995}, \emph{243}, 409--418\relax
\mciteBstWouldAddEndPuncttrue
\mciteSetBstMidEndSepPunct{\mcitedefaultmidpunct}
{\mcitedefaultendpunct}{\mcitedefaultseppunct}\relax
\EndOfBibitem
\bibitem[Salpeter and Bethe(1951)Salpeter, and Bethe]{Bet51}
Salpeter,~E.~E.; Bethe,~H.~A. A Relativistic Equation for Bound-State Problems.
  \emph{Phys. Rev.} \textbf{1951}, \emph{84}, 1232--1242\relax
\mciteBstWouldAddEndPuncttrue
\mciteSetBstMidEndSepPunct{\mcitedefaultmidpunct}
{\mcitedefaultendpunct}{\mcitedefaultseppunct}\relax
\EndOfBibitem
\bibitem[Strinati(1988)]{Str88}
Strinati,~G. Application of the {{Green}}'s Functions Method to the Study of
  the Optical Properties of Semiconductors. \emph{Riv. Nuovo Cimento}
  \textbf{1988}, \emph{11}, 1--86\relax
\mciteBstWouldAddEndPuncttrue
\mciteSetBstMidEndSepPunct{\mcitedefaultmidpunct}
{\mcitedefaultendpunct}{\mcitedefaultseppunct}\relax
\EndOfBibitem
\bibitem[Blase \latin{et~al.}(2018)Blase, Duchemin, and Jacquemin]{Bla18}
Blase,~X.; Duchemin,~I.; Jacquemin,~D. The Bethe-Salpeter Equation in
  Chemistry: Relations with TD-DFT{,} Applications and Challenges. \emph{Chem.
  Soc. Rev.} \textbf{2018}, \emph{47}, 1022--1043\relax
\mciteBstWouldAddEndPuncttrue
\mciteSetBstMidEndSepPunct{\mcitedefaultmidpunct}
{\mcitedefaultendpunct}{\mcitedefaultseppunct}\relax
\EndOfBibitem
\bibitem[Nooijen(1999)]{Noo99}
Nooijen,~M. Similarity Transformed Equation of Motion Coupled-Cluster Study of
  Excited States of Selected Azabenzenes. \emph{SpectroChim. Acta A}
  \textbf{1999}, \emph{55}, 539--559\relax
\mciteBstWouldAddEndPuncttrue
\mciteSetBstMidEndSepPunct{\mcitedefaultmidpunct}
{\mcitedefaultendpunct}{\mcitedefaultseppunct}\relax
\EndOfBibitem
\bibitem[Dutta \latin{et~al.}(2018)Dutta, Nooijen, Neese, and Izs{\'a}k]{Dut18}
Dutta,~A.~K.; Nooijen,~M.; Neese,~F.; Izs{\'a}k,~R. Exploring the Accuracy of a
  Low Scaling Similarity Transformed Equation of Motion Method for Vertical
  Excitation Energies. \emph{J. Chem. Theory Comput.} \textbf{2018}, \emph{14},
  72--91\relax
\mciteBstWouldAddEndPuncttrue
\mciteSetBstMidEndSepPunct{\mcitedefaultmidpunct}
{\mcitedefaultendpunct}{\mcitedefaultseppunct}\relax
\EndOfBibitem
\bibitem[Schreiber \latin{et~al.}(2008)Schreiber, Silva-Junior, Sauer, and
  Thiel]{Sch08}
Schreiber,~M.; Silva-Junior,~M.~R.; Sauer,~S. P.~A.; Thiel,~W. Benchmarks for
  Electronically Excited States: CASPT2, CC2, CCSD and CC3. \emph{J. Chem.
  Phys.} \textbf{2008}, \emph{128}, 134110\relax
\mciteBstWouldAddEndPuncttrue
\mciteSetBstMidEndSepPunct{\mcitedefaultmidpunct}
{\mcitedefaultendpunct}{\mcitedefaultseppunct}\relax
\EndOfBibitem
\bibitem[Sauer \latin{et~al.}(2009)Sauer, Schreiber, Silva-Junior, and
  Thiel]{Sau09}
Sauer,~S. P.~A.; Schreiber,~M.; Silva-Junior,~M.~R.; Thiel,~W. Benchmarks for
  Electronically Excited States: A Comparison of Noniterative and Iterative
  Triples Corrections in Linear Response Coupled Cluster Methods: CCSDR(3)
  \emph{versus} CC3. \emph{J. Chem. Theory Comput.} \textbf{2009}, \emph{5},
  555--564\relax
\mciteBstWouldAddEndPuncttrue
\mciteSetBstMidEndSepPunct{\mcitedefaultmidpunct}
{\mcitedefaultendpunct}{\mcitedefaultseppunct}\relax
\EndOfBibitem
\bibitem[Silva-Junior \latin{et~al.}(2010)Silva-Junior, Schreiber, Sauer, and
  Thiel]{Sil10c}
Silva-Junior,~M.~R.; Schreiber,~M.; Sauer,~S. P.~A.; Thiel,~W. Benchmarks of
  Electronically Excited States: Basis Set Effecs Benchmarks of Electronically
  Excited States: Basis Set Effects on CASPT2 Results. \emph{J. Chem. Phys.}
  \textbf{2010}, \emph{133}, 174318\relax
\mciteBstWouldAddEndPuncttrue
\mciteSetBstMidEndSepPunct{\mcitedefaultmidpunct}
{\mcitedefaultendpunct}{\mcitedefaultseppunct}\relax
\EndOfBibitem
\bibitem[Loos \latin{et~al.}(2018)Loos, Scemama, Blondel, Garniron, Caffarel,
  and Jacquemin]{Loo18a}
Loos,~P.-F.; Scemama,~A.; Blondel,~A.; Garniron,~Y.; Caffarel,~M.;
  Jacquemin,~D. A Mountaineering Strategy to Excited States: Highly-Accurate
  Reference Energies and Benchmarks. \emph{J. Chem. Theory Comput.}
  \textbf{2018}, \emph{14}, 4360--4379\relax
\mciteBstWouldAddEndPuncttrue
\mciteSetBstMidEndSepPunct{\mcitedefaultmidpunct}
{\mcitedefaultendpunct}{\mcitedefaultseppunct}\relax
\EndOfBibitem
\bibitem[Loos \latin{et~al.}(2019)Loos, Boggio-Pasqua, Scemama, Caffarel, and
  Jacquemin]{Loo19c}
Loos,~P.-F.; Boggio-Pasqua,~M.; Scemama,~A.; Caffarel,~M.; Jacquemin,~D.
  Reference Energies for Double Excitations. \emph{J. Chem. Theory Comput.}
  \textbf{2019}, \emph{15}, 1939--1956\relax
\mciteBstWouldAddEndPuncttrue
\mciteSetBstMidEndSepPunct{\mcitedefaultmidpunct}
{\mcitedefaultendpunct}{\mcitedefaultseppunct}\relax
\EndOfBibitem
\bibitem[Loos \latin{et~al.}(2020)Loos, Lipparini, Boggio-Pasqua, Scemama, and
  Jacquemin]{Loo20a}
Loos,~P.-F.; Lipparini,~F.; Boggio-Pasqua,~M.; Scemama,~A.; Jacquemin,~D. A
  Mountaineering Strategy to Excited States: Highly-Accurate Energies and
  Benchmarks for Medium Size Molecules. \emph{J. Chem. Theory Comput.}
  \textbf{2020}, \emph{16}, 1711--1741\relax
\mciteBstWouldAddEndPuncttrue
\mciteSetBstMidEndSepPunct{\mcitedefaultmidpunct}
{\mcitedefaultendpunct}{\mcitedefaultseppunct}\relax
\EndOfBibitem
\bibitem[Loos \latin{et~al.}(2020)Loos, Scemama, Boggio-Pasqua, and
  Jacquemin]{Loo20d}
Loos,~P.-F.; Scemama,~A.; Boggio-Pasqua,~M.; Jacquemin,~D. A Mountaineering
  Strategy to Excited States: Highly-Accurate Energies and Benchmarks for
  Exotic Molecules and Radicals. \emph{J. Chem. Theory Comput.} \textbf{2020},
  \emph{16}, 3720--3736\relax
\mciteBstWouldAddEndPuncttrue
\mciteSetBstMidEndSepPunct{\mcitedefaultmidpunct}
{\mcitedefaultendpunct}{\mcitedefaultseppunct}\relax
\EndOfBibitem
\bibitem[Dierksen and Grimme(2004)Dierksen, and Grimme]{Die04b}
Dierksen,~M.; Grimme,~S. The Vibronic Structure of Electronic Absorption
  Spectra of Large Molecules: A Time-Dependent Density Functional Study on the
  Influence of \emph{Exact} Hartree-Fock Exchange. \emph{J. Phys. Chem. A}
  \textbf{2004}, \emph{108}, 10225--10237\relax
\mciteBstWouldAddEndPuncttrue
\mciteSetBstMidEndSepPunct{\mcitedefaultmidpunct}
{\mcitedefaultendpunct}{\mcitedefaultseppunct}\relax
\EndOfBibitem
\bibitem[Send \latin{et~al.}(2011)Send, K{\"u}hn, and Furche]{Sen11b}
Send,~R.; K{\"u}hn,~M.; Furche,~F. Assessing Excited State Methods by Adiabatic
  Excitation Energies. \emph{J. Chem. Theory Comput.} \textbf{2011}, \emph{7},
  2376--2386\relax
\mciteBstWouldAddEndPuncttrue
\mciteSetBstMidEndSepPunct{\mcitedefaultmidpunct}
{\mcitedefaultendpunct}{\mcitedefaultseppunct}\relax
\EndOfBibitem
\bibitem[Loos and Jacquemin(2019)Loos, and Jacquemin]{Loo19a}
Loos,~P.-F.; Jacquemin,~D. Chemically Accurate 0-0 Energies with
  not-so-Accurate Excited State Geometries. \emph{J. Chem. Theory Comput.}
  \textbf{2019}, \emph{15}, 2481--2491\relax
\mciteBstWouldAddEndPuncttrue
\mciteSetBstMidEndSepPunct{\mcitedefaultmidpunct}
{\mcitedefaultendpunct}{\mcitedefaultseppunct}\relax
\EndOfBibitem
\bibitem[Loos and Jacquemin(2019)Loos, and Jacquemin]{Loo19b}
Loos,~P.-F.; Jacquemin,~D. Evaluating 0-0 Energies with Theoretical Tools: a
  Short Review. \emph{ChemPhotoChem} \textbf{2019}, \emph{3}, 684--696\relax
\mciteBstWouldAddEndPuncttrue
\mciteSetBstMidEndSepPunct{\mcitedefaultmidpunct}
{\mcitedefaultendpunct}{\mcitedefaultseppunct}\relax
\EndOfBibitem
\bibitem[Medvedev \latin{et~al.}(2017)Medvedev, Bushmarinov, Sun, Perdew, and
  Lyssenko]{Med17}
Medvedev,~M.~G.; Bushmarinov,~I.~S.; Sun,~J.; Perdew,~J.~P.; Lyssenko,~K.~A.
  Density Functional Theory is Straying From the Path Toward the Exact
  Functional. \emph{Science} \textbf{2017}, \emph{355}, 49--52\relax
\mciteBstWouldAddEndPuncttrue
\mciteSetBstMidEndSepPunct{\mcitedefaultmidpunct}
{\mcitedefaultendpunct}{\mcitedefaultseppunct}\relax
\EndOfBibitem
\bibitem[Hait and Head-Gordon(2018)Hait, and Head-Gordon]{Hai18}
Hait,~D.; Head-Gordon,~M. How Accurate Is Density Functional Theory at
  Predicting Dipole Moments? An Assessment Using a New Database of 200
  Benchmark Values. \emph{J. Chem. Theory Comput.} \textbf{2018}, \emph{14},
  1969--1981\relax
\mciteBstWouldAddEndPuncttrue
\mciteSetBstMidEndSepPunct{\mcitedefaultmidpunct}
{\mcitedefaultendpunct}{\mcitedefaultseppunct}\relax
\EndOfBibitem
\bibitem[Hait \latin{et~al.}(2020)Hait, Liang, and Head-Gordon]{Hai20}
Hait,~D.; Liang,~Y.~H.; Head-Gordon,~M. Too Big, Too Small or Just Right? A
  Benchmark Assessment of Density Functional Theory for Predicting ehe Spatial
  Extent of the Electron Density of Small Chemical Systems. \textbf{2020},
  arXiv:2011.12561\relax
\mciteBstWouldAddEndPuncttrue
\mciteSetBstMidEndSepPunct{\mcitedefaultmidpunct}
{\mcitedefaultendpunct}{\mcitedefaultseppunct}\relax
\EndOfBibitem
\bibitem[Chrayteh \latin{et~al.}(2021)Chrayteh, Blondel, Loos, and
  Jacquemin]{Chr21}
Chrayteh,~A.; Blondel,~A.; Loos,~P.-F.; Jacquemin,~D. A Mountaineering Strategy
  to Excited States: Highly-Accurate Oscillator Strengths and Dipole Moments of
  Small Molecules. \emph{J. Chem. Theory Comput.} \textbf{2021}, \emph{17},
  416--438\relax
\mciteBstWouldAddEndPuncttrue
\mciteSetBstMidEndSepPunct{\mcitedefaultmidpunct}
{\mcitedefaultendpunct}{\mcitedefaultseppunct}\relax
\EndOfBibitem
\bibitem[Giner \latin{et~al.}(2019)Giner, Scemama, Toulouse, and Loos]{Gin19}
Giner,~E.; Scemama,~A.; Toulouse,~J.; Loos,~P.~F. Chemically Accurate
  Excitation Energies With Small Basis Sets. \emph{J. Chem. Phys.}
  \textbf{2019}, \emph{151}, 144118\relax
\mciteBstWouldAddEndPuncttrue
\mciteSetBstMidEndSepPunct{\mcitedefaultmidpunct}
{\mcitedefaultendpunct}{\mcitedefaultseppunct}\relax
\EndOfBibitem
\bibitem[Thomas(1925)]{Tho25}
Thomas,~W. {\"U}ber die Zahl der Dispersionselektronen, die einem
  station{\"a}ren Zustande zugeordnet sind. \emph{Naturwissenschaften}
  \textbf{1925}, \emph{13}, 627--627\relax
\mciteBstWouldAddEndPuncttrue
\mciteSetBstMidEndSepPunct{\mcitedefaultmidpunct}
{\mcitedefaultendpunct}{\mcitedefaultseppunct}\relax
\EndOfBibitem
\bibitem[Reiche and Thomas(1925)Reiche, and Thomas]{Rei25}
Reiche,~F.; Thomas,~W. {\"U}ber die Zahl der Dispersionselektronen, die einem
  station{\"a}ren Zustand zugeordnet sind. \emph{Z. Phys.} \textbf{1925},
  \emph{34}, 510--525\relax
\mciteBstWouldAddEndPuncttrue
\mciteSetBstMidEndSepPunct{\mcitedefaultmidpunct}
{\mcitedefaultendpunct}{\mcitedefaultseppunct}\relax
\EndOfBibitem
\bibitem[Kuhn(1925)]{Kuh25}
Kuhn,~W. {\"U}ber die Gesamtst{\"a}rke der von einem Zustande ausgehenden
  Absorptionslinien. \emph{Z. Phys.} \textbf{1925}, \emph{33}, 408--412\relax
\mciteBstWouldAddEndPuncttrue
\mciteSetBstMidEndSepPunct{\mcitedefaultmidpunct}
{\mcitedefaultendpunct}{\mcitedefaultseppunct}\relax
\EndOfBibitem
\bibitem[A.Sauer \latin{et~al.}(2019)A.Sauer, R.Sabin, and Oddershede]{Sau19}
A.Sauer,~S.~P.; R.Sabin,~J.; Oddershede,~J. Calculation of Mean Excitation
  Energies. \emph{Adv. Quantum Chem.} \textbf{2019}, \emph{80}, 225--245\relax
\mciteBstWouldAddEndPuncttrue
\mciteSetBstMidEndSepPunct{\mcitedefaultmidpunct}
{\mcitedefaultendpunct}{\mcitedefaultseppunct}\relax
\EndOfBibitem
\bibitem[Paw{\l}owski \latin{et~al.}(2004)Paw{\l}owski, J{\o}rgensen, and
  H{\"a}ttig]{Paw04}
Paw{\l}owski,~F.; J{\o}rgensen,~P.; H{\"a}ttig,~C. Gauge Invariance of
  Oscillator Strengths in the Approximate Coupled Cluster Triples Model CC3.
  \emph{Chem. Phys. Lett.} \textbf{2004}, \emph{389}, 413--420\relax
\mciteBstWouldAddEndPuncttrue
\mciteSetBstMidEndSepPunct{\mcitedefaultmidpunct}
{\mcitedefaultendpunct}{\mcitedefaultseppunct}\relax
\EndOfBibitem
\bibitem[Pedersen and Koch(1998)Pedersen, and Koch]{Ped98}
Pedersen,~T.~B.; Koch,~H. Gauge Invariance of the Coupled Cluster Oscillator
  Strength. \emph{Chem. Phys. Lett.} \textbf{1998}, \emph{293}, 251--260\relax
\mciteBstWouldAddEndPuncttrue
\mciteSetBstMidEndSepPunct{\mcitedefaultmidpunct}
{\mcitedefaultendpunct}{\mcitedefaultseppunct}\relax
\EndOfBibitem
\bibitem[Helgaker \latin{et~al.}(1989)Helgaker, J{\o}rgensen, and Handy]{Hel89}
Helgaker,~T.; J{\o}rgensen,~P.; Handy,~N. A Numerically Stable Procedure for
  Calculating M{\o}ller-Plesset Energy Derivatives, Derived Using the Theory of
  Lagrangians. \emph{Theoret. Chim. Acta} \textbf{1989}, \emph{76},
  227--245\relax
\mciteBstWouldAddEndPuncttrue
\mciteSetBstMidEndSepPunct{\mcitedefaultmidpunct}
{\mcitedefaultendpunct}{\mcitedefaultseppunct}\relax
\EndOfBibitem
\bibitem[Koch \latin{et~al.}(1990)Koch, Jensen, Jorgensen, Helgaker, Scuseria,
  and Schaefer]{Koc90}
Koch,~H.; Jensen,~H. J.~A.; Jorgensen,~P.; Helgaker,~T.; Scuseria,~G.~E.;
  Schaefer,~H.~F. Coupled Cluster Energy Derivatives. Analytic Hessian for the
  Closed‐Shell Coupled Cluster Singles and Doubles Wave Function: Theory and
  Applications. \emph{J. Chem. Phys.} \textbf{1990}, \emph{92},
  4924--4940\relax
\mciteBstWouldAddEndPuncttrue
\mciteSetBstMidEndSepPunct{\mcitedefaultmidpunct}
{\mcitedefaultendpunct}{\mcitedefaultseppunct}\relax
\EndOfBibitem
\bibitem[H{\"a}ttig(2003)]{Hat03}
H{\"a}ttig,~C. Geometry Optimizations with the Coupled-Cluster Model CC2 Using
  the Resolution-of-the-Identity Approximation. \emph{J. Chem. Phys.}
  \textbf{2003}, \emph{118}, 7751--7761\relax
\mciteBstWouldAddEndPuncttrue
\mciteSetBstMidEndSepPunct{\mcitedefaultmidpunct}
{\mcitedefaultendpunct}{\mcitedefaultseppunct}\relax
\EndOfBibitem
\bibitem[H{\"a}ttig(2005)]{Hat05c}
H{\"a}ttig,~C. In \emph{Response Theory and Molecular Properties (A Tribute to
  Jan Linderberg and Poul J{\o}rgensen)}; Jensen,~H.~A., Ed.; Advances in
  Quantum Chemistry; Academic Press, 2005; Vol.~50; pp 37--60\relax
\mciteBstWouldAddEndPuncttrue
\mciteSetBstMidEndSepPunct{\mcitedefaultmidpunct}
{\mcitedefaultendpunct}{\mcitedefaultseppunct}\relax
\EndOfBibitem
\bibitem[Monkhorst(1977)]{Mon77}
Monkhorst,~H.~J. Calculation of Properties with the Coupled-Cluster Method.
  \emph{Int. J. Quantum Chem.} \textbf{1977}, \emph{12}, 421--432\relax
\mciteBstWouldAddEndPuncttrue
\mciteSetBstMidEndSepPunct{\mcitedefaultmidpunct}
{\mcitedefaultendpunct}{\mcitedefaultseppunct}\relax
\EndOfBibitem
\bibitem[Christiansen \latin{et~al.}(1998)Christiansen, J{\o}rgensen, and
  H\"attig]{Chr98d}
Christiansen,~O.; J{\o}rgensen,~P.; H\"attig,~C. Response Functions from
  Fourier Component Variational Perturbation Theory Applied to a Time-Averaged
  Quasienergy. \emph{Int. J. Quantum Chem.} \textbf{1998}, \emph{68},
  1--52\relax
\mciteBstWouldAddEndPuncttrue
\mciteSetBstMidEndSepPunct{\mcitedefaultmidpunct}
{\mcitedefaultendpunct}{\mcitedefaultseppunct}\relax
\EndOfBibitem
\bibitem[K{\'a}llay and Gauss(2004)K{\'a}llay, and Gauss]{Kal04}
K{\'a}llay,~M.; Gauss,~J. Calculation of Excited-State Properties Using General
  Coupled-Cluster and Configuration-Interaction Models. \emph{J. Chem. Phys.}
  \textbf{2004}, \emph{121}, 9257--9269\relax
\mciteBstWouldAddEndPuncttrue
\mciteSetBstMidEndSepPunct{\mcitedefaultmidpunct}
{\mcitedefaultendpunct}{\mcitedefaultseppunct}\relax
\EndOfBibitem
\bibitem[Salter \latin{et~al.}(1987)Salter, Sekino, and Bartlett]{Sal87}
Salter,~E.~A.; Sekino,~H.; Bartlett,~R.~J. Property Evaluation and Orbital
  Relaxation in Coupled Cluster Methods. \emph{J. Chem. Phys.} \textbf{1987},
  \emph{87}, 502--509\relax
\mciteBstWouldAddEndPuncttrue
\mciteSetBstMidEndSepPunct{\mcitedefaultmidpunct}
{\mcitedefaultendpunct}{\mcitedefaultseppunct}\relax
\EndOfBibitem
\bibitem[Trucks \latin{et~al.}(1988)Trucks, Salter, Sosa, and Bartlett]{Tru88}
Trucks,~G.~W.; Salter,~E.; Sosa,~C.; Bartlett,~R.~J. Theory and Implementation
  of the MBPT Density Matrix. An Application to One-Electron Properties.
  \emph{Chem. Phys. Lett.} \textbf{1988}, \emph{147}, 359--366\relax
\mciteBstWouldAddEndPuncttrue
\mciteSetBstMidEndSepPunct{\mcitedefaultmidpunct}
{\mcitedefaultendpunct}{\mcitedefaultseppunct}\relax
\EndOfBibitem
\bibitem[ROWE(1968)]{Row68}
ROWE,~D.~J. Equations-of-Motion Method and the Extended Shell Model. \emph{Rev.
  Mod. Phys.} \textbf{1968}, \emph{40}, 153--166\relax
\mciteBstWouldAddEndPuncttrue
\mciteSetBstMidEndSepPunct{\mcitedefaultmidpunct}
{\mcitedefaultendpunct}{\mcitedefaultseppunct}\relax
\EndOfBibitem
\bibitem[Stanton and Bartlett(1993)Stanton, and Bartlett]{Sta93}
Stanton,~J.~F.; Bartlett,~R.~J. The Equation of Motion Coupled-Cluster Method -
  A Systematic Biorthogonal Approach to Molecular Excitation Energies,
  Transition-Probabilities, and Excited-State Properties. \emph{J. Chem. Phys.}
  \textbf{1993}, \emph{98}, 7029--7039\relax
\mciteBstWouldAddEndPuncttrue
\mciteSetBstMidEndSepPunct{\mcitedefaultmidpunct}
{\mcitedefaultendpunct}{\mcitedefaultseppunct}\relax
\EndOfBibitem
\bibitem[Schirmer(1991)]{Sch91}
Schirmer,~J. Closed-Form Intermediate Representations of Many-Body Propagators
  and Resolvent Matrices. \emph{Phys. Rev. A.} \textbf{1991}, \emph{43},
  4647--4659\relax
\mciteBstWouldAddEndPuncttrue
\mciteSetBstMidEndSepPunct{\mcitedefaultmidpunct}
{\mcitedefaultendpunct}{\mcitedefaultseppunct}\relax
\EndOfBibitem
\bibitem[Schirmer and Mertins(2010)Schirmer, and Mertins]{Sch10}
Schirmer,~J.; Mertins,~F. Review of Biorthogonal Coupled Cluster
  Representations for Electronic Excitation. \emph{Theor. Chem. Acc.}
  \textbf{2010}, \emph{125}, 145--172\relax
\mciteBstWouldAddEndPuncttrue
\mciteSetBstMidEndSepPunct{\mcitedefaultmidpunct}
{\mcitedefaultendpunct}{\mcitedefaultseppunct}\relax
\EndOfBibitem
\bibitem[Hodecker \latin{et~al.}(2019)Hodecker, Rehn, Dreuw, and
  H{\"o}fener]{Hod19c}
Hodecker,~M.; Rehn,~D.~R.; Dreuw,~A.; H{\"o}fener,~S. Similarities and
  Differences of the Lagrange Formalism and the Intermediate State
  Representation in the Treatment of Molecular Properties. \emph{J. Chem.
  Phys.} \textbf{2019}, \emph{150}, 164125\relax
\mciteBstWouldAddEndPuncttrue
\mciteSetBstMidEndSepPunct{\mcitedefaultmidpunct}
{\mcitedefaultendpunct}{\mcitedefaultseppunct}\relax
\EndOfBibitem
\bibitem[Koch \latin{et~al.}(1994)Koch, Kobayashi, Sanchez~de Mer{\'a}s, and
  Jorgensen]{Koc94}
Koch,~H.; Kobayashi,~R.; Sanchez~de Mer{\'a}s,~A.; Jorgensen,~P. Calculation of
  Size‐Intensive Transition Moments from the Coupled Cluster Singles and
  Doubles Linear Response Function. \emph{J. Chem. Phys.} \textbf{1994},
  \emph{100}, 4393--4400\relax
\mciteBstWouldAddEndPuncttrue
\mciteSetBstMidEndSepPunct{\mcitedefaultmidpunct}
{\mcitedefaultendpunct}{\mcitedefaultseppunct}\relax
\EndOfBibitem
\bibitem[Caricato \latin{et~al.}(2009)Caricato, Trucks, and Frisch]{Car09b}
Caricato,~M.; Trucks,~G.~W.; Frisch,~M.~J. On the Difference Between the
  Transition Properties Calculated with Linear Response- and Equation of
  Motion-CCSD Approaches. \emph{J. Chem. Phys.} \textbf{2009}, \emph{131},
  174104\relax
\mciteBstWouldAddEndPuncttrue
\mciteSetBstMidEndSepPunct{\mcitedefaultmidpunct}
{\mcitedefaultendpunct}{\mcitedefaultseppunct}\relax
\EndOfBibitem
\bibitem[K{\'a}nn{\'a}r and Szalay(2014)K{\'a}nn{\'a}r, and Szalay]{Kan14}
K{\'a}nn{\'a}r,~D.; Szalay,~P.~G. Benchmarking Coupled Cluster Methods on
  Valence Singlet Excited States. \emph{J. Chem. Theory Comput.} \textbf{2014},
  \emph{10}, 3757--3765\relax
\mciteBstWouldAddEndPuncttrue
\mciteSetBstMidEndSepPunct{\mcitedefaultmidpunct}
{\mcitedefaultendpunct}{\mcitedefaultseppunct}\relax
\EndOfBibitem
\bibitem[Tawada \latin{et~al.}(2004)Tawada, Tsuneda, Yanagisawa, Yanai, and
  Hirao]{Taw04}
Tawada,~T.; Tsuneda,~T.; Yanagisawa,~S.; Yanai,~T.; Hirao,~K. A
  Long-Range-Corrected Time-Dependent Density Functional Theory. \emph{J. Chem.
  Phys.} \textbf{2004}, \emph{120}, 8425--8433\relax
\mciteBstWouldAddEndPuncttrue
\mciteSetBstMidEndSepPunct{\mcitedefaultmidpunct}
{\mcitedefaultendpunct}{\mcitedefaultseppunct}\relax
\EndOfBibitem
\bibitem[Miura \latin{et~al.}(2007)Miura, Aoki, and Champagne]{Miu07}
Miura,~M.; Aoki,~Y.; Champagne,~B. Assessment of Time-Dependent Density
  Functional Schemes for Computing The Oscillator Strengths of Benzene, Phenol,
  Aniline, and Fluorobenzene. \emph{J. Chem. Phys.} \textbf{2007}, \emph{127},
  084103\relax
\mciteBstWouldAddEndPuncttrue
\mciteSetBstMidEndSepPunct{\mcitedefaultmidpunct}
{\mcitedefaultendpunct}{\mcitedefaultseppunct}\relax
\EndOfBibitem
\bibitem[Timerghazin \latin{et~al.}(2008)Timerghazin, Carlson, Liang, Campbell,
  and Brown]{Tim08}
Timerghazin,~Q.~K.; Carlson,~H.~J.; Liang,~C.; Campbell,~R.~E.; Brown,~A.
  Computational Prediction of Absorbance Maxima for a Structurally Diverse
  Series of Engineered Green Fluorescent Protein Chromophores. \emph{J. Phys.
  Chem. B} \textbf{2008}, \emph{112}, 2533--2541\relax
\mciteBstWouldAddEndPuncttrue
\mciteSetBstMidEndSepPunct{\mcitedefaultmidpunct}
{\mcitedefaultendpunct}{\mcitedefaultseppunct}\relax
\EndOfBibitem
\bibitem[Silva-Junior \latin{et~al.}(2008)Silva-Junior, Schreiber, Sauer, and
  Thiel]{Sil08}
Silva-Junior,~M.~R.; Schreiber,~M.; Sauer,~S. P.~A.; Thiel,~W. Benchmarks for
  Electronically Excited States: Time-Dependent Density Functional Theory and
  Density Functional Theory Based Multireference Configuration Interaction.
  \emph{J. Chem. Phys.} \textbf{2008}, \emph{129}, 104103\relax
\mciteBstWouldAddEndPuncttrue
\mciteSetBstMidEndSepPunct{\mcitedefaultmidpunct}
{\mcitedefaultendpunct}{\mcitedefaultseppunct}\relax
\EndOfBibitem
\bibitem[Caricato \latin{et~al.}(2010)Caricato, Trucks, Frisch, and
  Wiberg]{Car10d}
Caricato,~M.; Trucks,~G.; Frisch,~M.; Wiberg,~K. Oscillator Strength: How Does
  TDDFT Compare to EOM-CCSD? \emph{J. Chem. Theory Comput.} \textbf{2010},
  \emph{7}, 456--466\relax
\mciteBstWouldAddEndPuncttrue
\mciteSetBstMidEndSepPunct{\mcitedefaultmidpunct}
{\mcitedefaultendpunct}{\mcitedefaultseppunct}\relax
\EndOfBibitem
\bibitem[Szalay \latin{et~al.}(2012)Szalay, Watson, Perera, Lotrich, and
  Bartlett]{Sza12}
Szalay,~P.~G.; Watson,~T.; Perera,~A.; Lotrich,~V.~F.; Bartlett,~R.~J.
  Benchmark Studies on the Building Blocks of DNA. 1. Superiority of Coupled
  Cluster Methods in Describing the Excited States of Nucleobases in the
  Franck--Condon Region. \emph{J. Phys. Chem. A} \textbf{2012}, \emph{116},
  6702--6710\relax
\mciteBstWouldAddEndPuncttrue
\mciteSetBstMidEndSepPunct{\mcitedefaultmidpunct}
{\mcitedefaultendpunct}{\mcitedefaultseppunct}\relax
\EndOfBibitem
\bibitem[Sauer \latin{et~al.}(2015)Sauer, Pitzner-Frydendahl, Buse, Jensen, and
  Thiel]{Sau15}
Sauer,~S.~P.; Pitzner-Frydendahl,~H.~F.; Buse,~M.; Jensen,~H. J.~A.; Thiel,~W.
  Performance of SOPPA-Based Methods in the Calculation of Vertical Excitation
  Energies and Oscillator Strengths. \emph{Mol. Phys.} \textbf{2015},
  \emph{113}, 2026--2045\relax
\mciteBstWouldAddEndPuncttrue
\mciteSetBstMidEndSepPunct{\mcitedefaultmidpunct}
{\mcitedefaultendpunct}{\mcitedefaultseppunct}\relax
\EndOfBibitem
\bibitem[Jacquemin \latin{et~al.}(2016)Jacquemin, Duchemin, Blondel, and
  Blase]{Jac16b}
Jacquemin,~D.; Duchemin,~I.; Blondel,~A.; Blase,~X. Assessment of the Accuracy
  of the Bethe-Salpeter (BSE/GW) Oscillator Strengths. \emph{J. Chem. Theory
  Comput.} \textbf{2016}, \emph{12}, 3969--3981\relax
\mciteBstWouldAddEndPuncttrue
\mciteSetBstMidEndSepPunct{\mcitedefaultmidpunct}
{\mcitedefaultendpunct}{\mcitedefaultseppunct}\relax
\EndOfBibitem
\bibitem[Robinson(2018)]{Rob18}
Robinson,~D. Comparison of the Transition Dipole Moments Calculated by TDDFT
  with High Level Wave Function Theory. \emph{J. Chem. Theory Comput.}
  \textbf{2018}, \emph{14}, 5303--5309\relax
\mciteBstWouldAddEndPuncttrue
\mciteSetBstMidEndSepPunct{\mcitedefaultmidpunct}
{\mcitedefaultendpunct}{\mcitedefaultseppunct}\relax
\EndOfBibitem
\bibitem[Silva-Junior \latin{et~al.}(2010)Silva-Junior, Sauer, Schreiber, and
  Thiel]{Sil10b}
Silva-Junior,~M.~R.; Sauer,~S. P.~A.; Schreiber,~M.; Thiel,~W. Basis Set
  Effects on Coupled Cluster Benchmarks of Electronically Excited States: CC3,
  CCSDR(3) and CC2. \emph{Mol. Phys.} \textbf{2010}, \emph{108}, 453--465\relax
\mciteBstWouldAddEndPuncttrue
\mciteSetBstMidEndSepPunct{\mcitedefaultmidpunct}
{\mcitedefaultendpunct}{\mcitedefaultseppunct}\relax
\EndOfBibitem
\bibitem[Furche and Ahlrichs(2002)Furche, and Ahlrichs]{Fur02}
Furche,~F.; Ahlrichs,~R. Adiabatic Time-Dependent Density Functional Methods
  for Excited States Properties. \emph{J. Chem. Phys.} \textbf{2002},
  \emph{117}, 7433--7447\relax
\mciteBstWouldAddEndPuncttrue
\mciteSetBstMidEndSepPunct{\mcitedefaultmidpunct}
{\mcitedefaultendpunct}{\mcitedefaultseppunct}\relax
\EndOfBibitem
\bibitem[King(2008)]{Kin08}
King,~R.~A. On the Accuracy of Computed Excited-State Dipole Moments. \emph{J.
  Phys. Chem. A} \textbf{2008}, \emph{112}, 5727--5733\relax
\mciteBstWouldAddEndPuncttrue
\mciteSetBstMidEndSepPunct{\mcitedefaultmidpunct}
{\mcitedefaultendpunct}{\mcitedefaultseppunct}\relax
\EndOfBibitem
\bibitem[Wong \latin{et~al.}(2009)Wong, Piacenza, and Della~Sala]{Won09b}
Wong,~B.~M.; Piacenza,~M.; Della~Sala,~F. Absorption and Fluorescence
  Properties of Oligothiophene Biomarkers from Long-Range-Corrected Absorption
  and Fluorescence Properties of Oligothiophene Biomarkersfrom
  Long-Range-Corrected Time-Dependent Density Functional Theory. \emph{Phys.
  Chem. Chem. Phys.} \textbf{2009}, \emph{11}, 4498--4508\relax
\mciteBstWouldAddEndPuncttrue
\mciteSetBstMidEndSepPunct{\mcitedefaultmidpunct}
{\mcitedefaultendpunct}{\mcitedefaultseppunct}\relax
\EndOfBibitem
\bibitem[Tapavicza \latin{et~al.}(2009)Tapavicza, Tavernelli, and
  Rothlisberger]{Tap09}
Tapavicza,~E.; Tavernelli,~I.; Rothlisberger,~U. Ab Initio Excited State
  Properties and Dynamics of a Prototype {$\sigma$}-Bridged-Donor-Acceptor
  Molecule. \emph{J. Phys. Chem. A} \textbf{2009}, \emph{113}, 9595--9602\relax
\mciteBstWouldAddEndPuncttrue
\mciteSetBstMidEndSepPunct{\mcitedefaultmidpunct}
{\mcitedefaultendpunct}{\mcitedefaultseppunct}\relax
\EndOfBibitem
\bibitem[Guido \latin{et~al.}(2010)Guido, Jacquemin, Adamo, and
  Mennucci]{Gui10}
Guido,~C.~A.; Jacquemin,~D.; Adamo,~C.; Mennucci,~B. On the TD-DFT Accuracy in
  Determining Single and Double Bonds in Excited-State Structures of Organic
  Molecules. \emph{J. Phys. Chem. A} \textbf{2010}, \emph{114},
  13402--13410\relax
\mciteBstWouldAddEndPuncttrue
\mciteSetBstMidEndSepPunct{\mcitedefaultmidpunct}
{\mcitedefaultendpunct}{\mcitedefaultseppunct}\relax
\EndOfBibitem
\bibitem[Silva-Junior and Thiel(2010)Silva-Junior, and Thiel]{Sil10}
Silva-Junior,~M.~R.; Thiel,~W. Benchmark of Electronically Excited States for
  Semiempirical Methods: MNDO, AM1, PM3, OM1, OM2, OM3, INDO/S, and INDO/S2.
  \emph{J. Chem. Theory Comput.} \textbf{2010}, \emph{6}, 1546--1564\relax
\mciteBstWouldAddEndPuncttrue
\mciteSetBstMidEndSepPunct{\mcitedefaultmidpunct}
{\mcitedefaultendpunct}{\mcitedefaultseppunct}\relax
\EndOfBibitem
\bibitem[Hellweg(2011)]{Hel11}
Hellweg,~A. The Accuracy of Dipole Moments from Spin-Component Scaled CC2 in
  Ground and Electronically Excited States. \emph{J. Chem. Phys.}
  \textbf{2011}, \emph{134}, 064103\relax
\mciteBstWouldAddEndPuncttrue
\mciteSetBstMidEndSepPunct{\mcitedefaultmidpunct}
{\mcitedefaultendpunct}{\mcitedefaultseppunct}\relax
\EndOfBibitem
\bibitem[Jacquemin(2016)]{Jac16d}
Jacquemin,~D. Excited-State Dipole and Quadrupole Moments: TD-DFT versus CC2.
  \emph{J. Chem. Theory Comput.} \textbf{2016}, \emph{12}, 3993--4003\relax
\mciteBstWouldAddEndPuncttrue
\mciteSetBstMidEndSepPunct{\mcitedefaultmidpunct}
{\mcitedefaultendpunct}{\mcitedefaultseppunct}\relax
\EndOfBibitem
\bibitem[Hodecker and Dreuw(2020)Hodecker, and Dreuw]{Hod20}
Hodecker,~M.; Dreuw,~A. Unitary Coupled Cluster Ground- and Excited-State
  Molecular Properties. \emph{J. Chem. Phys.} \textbf{2020}, \emph{153},
  084112\relax
\mciteBstWouldAddEndPuncttrue
\mciteSetBstMidEndSepPunct{\mcitedefaultmidpunct}
{\mcitedefaultendpunct}{\mcitedefaultseppunct}\relax
\EndOfBibitem
\bibitem[Christiansen \latin{et~al.}(1995)Christiansen, Koch, and
  J{\o}rgensen]{Chr95b}
Christiansen,~O.; Koch,~H.; J{\o}rgensen,~P. Response Functions in the CC3
  Iterative Triple Excitation Model. \emph{J. Chem. Phys.} \textbf{1995},
  \emph{103}, 7429--7441\relax
\mciteBstWouldAddEndPuncttrue
\mciteSetBstMidEndSepPunct{\mcitedefaultmidpunct}
{\mcitedefaultendpunct}{\mcitedefaultseppunct}\relax
\EndOfBibitem
\bibitem[Koch \latin{et~al.}(1995)Koch, Christiansen, Jorgensen, and
  Olsen]{Koc95}
Koch,~H.; Christiansen,~O.; Jorgensen,~P.; Olsen,~J. Excitation Energies of BH,
  CH$_2$ and Ne in Full Configuration Interaction and the Hierarchy CCS, CC2,
  CCSD and CC3 of Coupled Cluster Models. \emph{Chem. Phys. Lett.}
  \textbf{1995}, \emph{244}, 75--82\relax
\mciteBstWouldAddEndPuncttrue
\mciteSetBstMidEndSepPunct{\mcitedefaultmidpunct}
{\mcitedefaultendpunct}{\mcitedefaultseppunct}\relax
\EndOfBibitem
\bibitem[Aidas \latin{et~al.}(2014)Aidas, Angeli, Bak, Bakken, Bast, Boman,
  Christiansen, Cimiraglia, Coriani, Dahle, Dalskov, Ekstr{\"o}m, Enevoldsen,
  Eriksen, Ettenhuber, Fern{\'a}ndez, Ferrighi, Fliegl, Frediani, Hald,
  Halkier, H\"attig, Heiberg, Helgaker, Hennum, Hettema, Hjerten{\ae}s,
  H{\o}st, Hoyvik, Iozzi, Jansik, Jensen, Jonsson, J{\o}rgensen, Kauczor,
  Kirpekar, Kj{\ae}rgaard, Klopper, Knecht, Kobayashi, Koch, Kongsted, Krapp,
  Kristensen, Ligabue, Lutn{\ae}s, Melo, Mikkelsen, Myhre, Neiss, Nielsen,
  Norman, Olsen, Olsen, Osted, Packer, Pawlowski, Pedersen, Provasi, Reine,
  Rinkevicius, Ruden, Ruud, Rybkin, Sa{\l}ek, Samson, de~Mer{\'a}s, Saue,
  Sauer, Schimmelpfennig, Sneskov, Steindal, Sylvester-Hvid, Taylor, Teale,
  Tellgren, Tew, Thorvaldsen, Thogersen, Vahtras, Watson, Wilson, Ziolkowski,
  and {\AA}gren]{dalton}
Aidas,~K.; Angeli,~C.; Bak,~K.~L.; Bakken,~V.; Bast,~R.; Boman,~L.;
  Christiansen,~O.; Cimiraglia,~R.; Coriani,~S.; Dahle,~P.; Dalskov,~E.~K.;
  Ekstr{\"o}m,~U.; Enevoldsen,~T.; Eriksen,~J.~J.; Ettenhuber,~P.;
  Fern{\'a}ndez,~B.; Ferrighi,~L.; Fliegl,~H.; Frediani,~L.; Hald,~K.;
  Halkier,~A.; H\"attig,~C.; Heiberg,~H.; Helgaker,~T.; Hennum,~A.~C.;
  Hettema,~H.; Hjerten{\ae}s,~E.; H{\o}st,~S.; Hoyvik,~I.-M.; Iozzi,~M.~F.;
  Jansik,~B.; Jensen,~H. J.~A.; Jonsson,~D.; J{\o}rgensen,~P.; Kauczor,~J.;
  Kirpekar,~S.; Kj{\ae}rgaard,~T.; Klopper,~W.; Knecht,~S.; Kobayashi,~R.;
  Koch,~H.; Kongsted,~J.; Krapp,~A.; Kristensen,~K.; Ligabue,~A.;
  Lutn{\ae}s,~O.~B.; Melo,~J.~I.; Mikkelsen,~K.~V.; Myhre,~R.~H.; Neiss,~C.;
  Nielsen,~C.~B.; Norman,~P.; Olsen,~J.; Olsen,~J. M.~H.; Osted,~A.;
  Packer,~M.~J.; Pawlowski,~F.; Pedersen,~T.~B.; Provasi,~P.~F.; Reine,~S.;
  Rinkevicius,~Z.; Ruden,~T.~A.; Ruud,~K.; Rybkin,~V.~V.; Sa{\l}ek,~P.;
  Samson,~C. C.~M.; de~Mer{\'a}s,~A.~S.; Saue,~T.; Sauer,~S. P.~A.;
  Schimmelpfennig,~B.; Sneskov,~K.; Steindal,~A.~H.; Sylvester-Hvid,~K.~O.;
  Taylor,~P.~R.; Teale,~A.~M.; Tellgren,~E.~I.; Tew,~D.~P.; Thorvaldsen,~A.~J.;
  Thogersen,~L.; Vahtras,~O.; Watson,~M.~A.; Wilson,~D. J.~D.; Ziolkowski,~M.;
  {\AA}gren,~H. The Dalton Quantum Chemistry Program System. \emph{WIREs
  Comput. Mol. Sci.} \textbf{2014}, \emph{4}, 269--284\relax
\mciteBstWouldAddEndPuncttrue
\mciteSetBstMidEndSepPunct{\mcitedefaultmidpunct}
{\mcitedefaultendpunct}{\mcitedefaultseppunct}\relax
\EndOfBibitem
\bibitem[cfo()]{cfour}
CFOUR, Coupled-Cluster Techniques for Computational Chemistry, a
  Quantum-Chemical Program Package by J.F. Stanton, J. Gauss, L. Cheng, M.E.
  Harding, D.A. Matthews, P.G. Szalay with contributions from A.A. Auer, R.J.
  Bartlett, U. Benedikt, C. Berger, D.E. Bernholdt, Y.J. Bomble, O.
  Christiansen, F. Engel, R. Faber, M. Heckert, O. Heun, M. Hilgenberg, C.
  Huber, T.-C. Jagau, D. Jonsson, J. Jus{\'e}lius, T. Kirsch, K. Klein, W.J.
  Lauderdale, F. Lipparini, T. Metzroth, L.A. M{\"u}ck, D.P. O'Neill, D.R.
  Price, E. Prochnow, C. Puzzarini, K. Ruud, F. Schiffmann, W. Schwalbach, C.
  Simmons, S. Stopkowicz, A. Tajti, J. V{\'a}zquez, F. Wang, J.D. Watts and the
  integral packages MOLECULE (J. Alml\"of and P.R. Taylor), PROPS (P.R.
  Taylor), ABACUS (T. Helgaker, H.J. Aa. Jensen, P. J{\o}rgensen, and J.
  Olsen), and ECP routines by A. V. Mitin and C. van W{\"u}llen. For the
  current version, see http://www.cfour.de.\relax
\mciteBstWouldAddEndPunctfalse
\mciteSetBstMidEndSepPunct{\mcitedefaultmidpunct}
{}{\mcitedefaultseppunct}\relax
\EndOfBibitem
\bibitem[Hattig \latin{et~al.}(2012)Hattig, Klopper, Kohn, and Tew]{Hat12}
Hattig,~C.; Klopper,~W.; Kohn,~A.; Tew,~D.~P. Explicitly Correlated Electrons
  in Molecules. \emph{Chem. Rev.} \textbf{2012}, \emph{112}, 4--74\relax
\mciteBstWouldAddEndPuncttrue
\mciteSetBstMidEndSepPunct{\mcitedefaultmidpunct}
{\mcitedefaultendpunct}{\mcitedefaultseppunct}\relax
\EndOfBibitem
\bibitem[Loos \latin{et~al.}(2019)Loos, Pradines, Scemama, Toulouse, and
  Giner]{Loo19}
Loos,~P.~F.; Pradines,~B.; Scemama,~A.; Toulouse,~J.; Giner,~E. A Density-Based
  Basis-Set Correction for Wave Function Theory. \emph{J. Phys. Chem. Lett.}
  \textbf{2019}, \emph{10}, 2931--2937\relax
\mciteBstWouldAddEndPuncttrue
\mciteSetBstMidEndSepPunct{\mcitedefaultmidpunct}
{\mcitedefaultendpunct}{\mcitedefaultseppunct}\relax
\EndOfBibitem
\bibitem[Kallay \latin{et~al.}(2020)Kallay, Nagy, Mester, Rolik, Samu, Csoka,
  Szabo, Gyevi-Nagu, Hegely, Ladjanski, Szegedy, Ladoczki, Petrov, Farkas,
  Mezei, and Ganyecz]{Kal20}
Kallay,~M.; Nagy,~P.~E.; Mester,~D.; Rolik,~Z.; Samu,~J.,~G. Ans~Csontos;
  Csoka,~J.; Szabo,~P.~B.; Gyevi-Nagu,~L.; Hegely,~B.; Ladjanski,~I.;
  Szegedy,~L.; Ladoczki,~B.; Petrov,~K.; Farkas,~M.; Mezei,~P.~D.; Ganyecz,~A.
  The MRCC Program System: Accurate Quantum Chemistry from Water to Proteins.
  \emph{J. Chem. Phys.} \textbf{2020}, \emph{152}, 074107\relax
\mciteBstWouldAddEndPuncttrue
\mciteSetBstMidEndSepPunct{\mcitedefaultmidpunct}
{\mcitedefaultendpunct}{\mcitedefaultseppunct}\relax
\EndOfBibitem
\bibitem[K{\'a}llay \latin{et~al.}(2017)K{\'a}llay, Rolik, Csontos, Nagy, Samu,
  Mester, Cs{\'o}ka, Szab{\'o}, Ladj{\'a}nszki, Szegedy, Lad{\'o}czki, Petrov,
  Farkas, Mezei, and H{\'e}gely.]{mrcc}
K{\'a}llay,~M.; Rolik,~Z.; Csontos,~J.; Nagy,~P.; Samu,~G.; Mester,~D.;
  Cs{\'o}ka,~J.; Szab{\'o},~B.; Ladj{\'a}nszki,~I.; Szegedy,~L.;
  Lad{\'o}czki,~B.; Petrov,~K.; Farkas,~M.; Mezei,~P.~D.; H{\'e}gely.,~B. MRCC,
  Quantum Chemical Program. 2017; See: www.mrcc.hu.\relax
\mciteBstWouldAddEndPunctfalse
\mciteSetBstMidEndSepPunct{\mcitedefaultmidpunct}
{}{\mcitedefaultseppunct}\relax
\EndOfBibitem
\bibitem[Purvis~III and Bartlett(1982)Purvis~III, and Bartlett]{Pur82}
Purvis~III,~G.~P.; Bartlett,~R.~J. A Full Coupled-Cluster Singles and Doubles
  Model: The Inclusion of Disconnected Triples. \emph{J. Chem. Phys.}
  \textbf{1982}, \emph{76}, 1910--1918\relax
\mciteBstWouldAddEndPuncttrue
\mciteSetBstMidEndSepPunct{\mcitedefaultmidpunct}
{\mcitedefaultendpunct}{\mcitedefaultseppunct}\relax
\EndOfBibitem
\bibitem[Scuseria \latin{et~al.}(1987)Scuseria, Scheiner, Lee, Rice, and
  Schaefer]{Scu87}
Scuseria,~G.~E.; Scheiner,~A.~C.; Lee,~T.~J.; Rice,~J.~E.; Schaefer,~H.~F. The
  Closed-Shell Coupled Cluster Single and Double Excitation (CCSD) Model for
  the Description of Electron Correlation. A Comparison with Configuration
  Interaction (CISD) Results. \emph{J. Chem. Phys.} \textbf{1987}, \emph{86},
  2881--2890\relax
\mciteBstWouldAddEndPuncttrue
\mciteSetBstMidEndSepPunct{\mcitedefaultmidpunct}
{\mcitedefaultendpunct}{\mcitedefaultseppunct}\relax
\EndOfBibitem
\bibitem[Koch \latin{et~al.}(1990)Koch, Jensen, Jorgensen, and
  Helgaker]{Koc90b}
Koch,~H.; Jensen,~H. J.~A.; Jorgensen,~P.; Helgaker,~T. Excitation Energies
  from the Coupled Cluster Singles and Doubles Linear Response Function
  (CCSDLR). Applications to Be, CH$^+$, CO, and H$_2$O. \emph{J. Chem. Phys.}
  \textbf{1990}, \emph{93}, 3345--3350\relax
\mciteBstWouldAddEndPuncttrue
\mciteSetBstMidEndSepPunct{\mcitedefaultmidpunct}
{\mcitedefaultendpunct}{\mcitedefaultseppunct}\relax
\EndOfBibitem
\bibitem[Stanton(1993)]{Sta93b}
Stanton,~J.~F. Many-Body Methods for Excited State Potential Energy Surfaces.
  I: General Theory of Energy Gradients for the Equation-of-Motion
  Coupled-Cluster Method. \emph{J. Chem. Phys.} \textbf{1993}, \emph{99},
  8840--8847\relax
\mciteBstWouldAddEndPuncttrue
\mciteSetBstMidEndSepPunct{\mcitedefaultmidpunct}
{\mcitedefaultendpunct}{\mcitedefaultseppunct}\relax
\EndOfBibitem
\bibitem[Noga and Bartlett(1987)Noga, and Bartlett]{Nog87}
Noga,~J.; Bartlett,~R.~J. The Full CCSDT Model for Molecular Electronic
  Structure. \emph{J. Chem. Phys.} \textbf{1987}, \emph{86}, 7041--7050\relax
\mciteBstWouldAddEndPuncttrue
\mciteSetBstMidEndSepPunct{\mcitedefaultmidpunct}
{\mcitedefaultendpunct}{\mcitedefaultseppunct}\relax
\EndOfBibitem
\bibitem[Scuseria and Schaefer(1988)Scuseria, and Schaefer]{Scu88}
Scuseria,~G.~E.; Schaefer,~H.~F. A New Implementation of the Full CCSDT Model
  for Molecular Electronic Structure. \emph{Chem. Phys. Lett.} \textbf{1988},
  \emph{152}, 382--386\relax
\mciteBstWouldAddEndPuncttrue
\mciteSetBstMidEndSepPunct{\mcitedefaultmidpunct}
{\mcitedefaultendpunct}{\mcitedefaultseppunct}\relax
\EndOfBibitem
\bibitem[Kucharski \latin{et~al.}(2001)Kucharski, W{\l}och, Musia{\l}, and
  Bartlett]{Kuc01}
Kucharski,~S.~A.; W{\l}och,~M.; Musia{\l},~M.; Bartlett,~R.~J. Coupled-Cluster
  Theory for Excited Electronic States: The Full Equation-Of-Motion
  Coupled-Cluster Single, Double, and Triple Excitation Method. \emph{J. Chem.
  Phys.} \textbf{2001}, \emph{115}, 8263--8266\relax
\mciteBstWouldAddEndPuncttrue
\mciteSetBstMidEndSepPunct{\mcitedefaultmidpunct}
{\mcitedefaultendpunct}{\mcitedefaultseppunct}\relax
\EndOfBibitem
\bibitem[Kowalski and Piecuch(2001)Kowalski, and Piecuch]{Kow01}
Kowalski,~K.; Piecuch,~P. The Active-Space Equation-of-Motion Coupled-Cluster
  Methods for Excited Electronic States: Full EOMCCSDt. \emph{J. Chem. Phys.}
  \textbf{2001}, \emph{115}, 643--651\relax
\mciteBstWouldAddEndPuncttrue
\mciteSetBstMidEndSepPunct{\mcitedefaultmidpunct}
{\mcitedefaultendpunct}{\mcitedefaultseppunct}\relax
\EndOfBibitem
\bibitem[Kowalski and Piecuch(2001)Kowalski, and Piecuch]{Kow01b}
Kowalski,~K.; Piecuch,~P. Excited-State Potential Energy Curves of CH$^+$: a
  Comparison of the EOMCCSDt And Full EOMCCSDT Results. \emph{Chem. Phys.
  Lett.} \textbf{2001}, \emph{347}, 237--246\relax
\mciteBstWouldAddEndPuncttrue
\mciteSetBstMidEndSepPunct{\mcitedefaultmidpunct}
{\mcitedefaultendpunct}{\mcitedefaultseppunct}\relax
\EndOfBibitem
\bibitem[Kucharski and Bartlett(1991)Kucharski, and Bartlett]{Kuc91}
Kucharski,~S.~A.; Bartlett,~R.~J. Recursive Intermediate Factorization and
  Complete Computational Linearization of the Coupled-Cluster Single, Double,
  Triple, and Quadruple Excitation Equations. \emph{Theor. Chim. Acta}
  \textbf{1991}, \emph{80}, 387--405\relax
\mciteBstWouldAddEndPuncttrue
\mciteSetBstMidEndSepPunct{\mcitedefaultmidpunct}
{\mcitedefaultendpunct}{\mcitedefaultseppunct}\relax
\EndOfBibitem
\bibitem[K{\'a}llay \latin{et~al.}(2003)K{\'a}llay, Gauss, and Szalay]{Kal03}
K{\'a}llay,~M.; Gauss,~J.; Szalay,~P.~G. Analytic First Derivatives for General
  Coupled-Cluster and Configuration Interaction Models. \emph{J. Chem. Phys.}
  \textbf{2003}, \emph{119}, 2991--3004\relax
\mciteBstWouldAddEndPuncttrue
\mciteSetBstMidEndSepPunct{\mcitedefaultmidpunct}
{\mcitedefaultendpunct}{\mcitedefaultseppunct}\relax
\EndOfBibitem
\bibitem[Hirata(2004)]{Hir04}
Hirata,~S. Higher-Order Equation-of-Motion Coupled-Cluster Methods. \emph{J.
  Chem. Phys.} \textbf{2004}, \emph{121}, 51--59\relax
\mciteBstWouldAddEndPuncttrue
\mciteSetBstMidEndSepPunct{\mcitedefaultmidpunct}
{\mcitedefaultendpunct}{\mcitedefaultseppunct}\relax
\EndOfBibitem
\bibitem[Balabanov and Peterson(2006)Balabanov, and Peterson]{Bal06}
Balabanov,~N.~B.; Peterson,~K.~A. Basis set Limit Electronic Excitation
  Energies, Ionization Potentials, and Electron Affinities for the $3d$
  Transition Metal Atoms: Coupled Cluster and Multireference Methods. \emph{J.
  Chem. Phys.} \textbf{2006}, \emph{125}, 074110\relax
\mciteBstWouldAddEndPuncttrue
\mciteSetBstMidEndSepPunct{\mcitedefaultmidpunct}
{\mcitedefaultendpunct}{\mcitedefaultseppunct}\relax
\EndOfBibitem
\bibitem[Kamiya and Hirata(2006)Kamiya, and Hirata]{Kam06b}
Kamiya,~M.; Hirata,~S. Higher-Order Equation-of-Motion Coupled-Cluster Methods
  for Ionization Processes. \emph{J. Chem. Phys.} \textbf{2006}, \emph{125},
  074111\relax
\mciteBstWouldAddEndPuncttrue
\mciteSetBstMidEndSepPunct{\mcitedefaultmidpunct}
{\mcitedefaultendpunct}{\mcitedefaultseppunct}\relax
\EndOfBibitem
\bibitem[Watson and Chan(2012)Watson, and Chan]{Wat12}
Watson,~M.~A.; Chan,~G. K.-L. Excited States of Butadiene to Chemical Accuracy:
  Reconciling Theory and Experiment. \emph{J. Chem. Theory Comput.}
  \textbf{2012}, \emph{8}, 4013--4018\relax
\mciteBstWouldAddEndPuncttrue
\mciteSetBstMidEndSepPunct{\mcitedefaultmidpunct}
{\mcitedefaultendpunct}{\mcitedefaultseppunct}\relax
\EndOfBibitem
\bibitem[Feller \latin{et~al.}(2014)Feller, Peterson, and Davidson]{Fel14}
Feller,~D.; Peterson,~K.~A.; Davidson,~E.~R. A Systematic Approach to
  Vertically Excited States of Ethylene Using Configuration Interaction and
  Coupled Cluster Techniques. \emph{J. Chem. Phys.} \textbf{2014}, \emph{141},
  104302\relax
\mciteBstWouldAddEndPuncttrue
\mciteSetBstMidEndSepPunct{\mcitedefaultmidpunct}
{\mcitedefaultendpunct}{\mcitedefaultseppunct}\relax
\EndOfBibitem
\bibitem[Franke \latin{et~al.}(2019)Franke, Moore, Schaefer, and
  Douberly]{Fra19}
Franke,~P.~R.; Moore,~K.~B.; Schaefer,~H.~F.; Douberly,~G.~E. tert-Butyl Peroxy
  Radical: Ground and First Excited State Energetics and Fundamental
  Frequencies. \emph{Phys. Chem. Chem. Phys.} \textbf{2019}, \emph{21},
  9747--9758\relax
\mciteBstWouldAddEndPuncttrue
\mciteSetBstMidEndSepPunct{\mcitedefaultmidpunct}
{\mcitedefaultendpunct}{\mcitedefaultseppunct}\relax
\EndOfBibitem
\bibitem[Andersson \latin{et~al.}(1990)Andersson, Malmqvist, Roos, Sadlej, and
  Wolinski]{And90}
Andersson,~K.; Malmqvist,~P.~A.; Roos,~B.~O.; Sadlej,~A.~J.; Wolinski,~K.
  Second-Order Perturbation Theory With a CASSCF Reference Function. \emph{J.
  Phys. Chem.} \textbf{1990}, \emph{94}, 5483--5488\relax
\mciteBstWouldAddEndPuncttrue
\mciteSetBstMidEndSepPunct{\mcitedefaultmidpunct}
{\mcitedefaultendpunct}{\mcitedefaultseppunct}\relax
\EndOfBibitem
\bibitem[Andersson \latin{et~al.}(1992)Andersson, Malmqvist, and Roos]{And92}
Andersson,~K.; Malmqvist,~P.-A.; Roos,~B.~O. Second-Order Perturbation Theory
  With a Complete Active Space Self-Consistent Field Reference Function.
  \emph{J. Chem. Phys.} \textbf{1992}, \emph{96}, 1218--1226\relax
\mciteBstWouldAddEndPuncttrue
\mciteSetBstMidEndSepPunct{\mcitedefaultmidpunct}
{\mcitedefaultendpunct}{\mcitedefaultseppunct}\relax
\EndOfBibitem
\bibitem[Roos and Andersson(1995)Roos, and Andersson]{Roo95}
Roos,~B.~O.; Andersson,~K. Multiconfigurational Perturbation Theory with Level
  Shift --- the Cr$_2$ Potential Revisited. \emph{Chem. Phys. Lett.}
  \textbf{1995}, \emph{245}, 215--223\relax
\mciteBstWouldAddEndPuncttrue
\mciteSetBstMidEndSepPunct{\mcitedefaultmidpunct}
{\mcitedefaultendpunct}{\mcitedefaultseppunct}\relax
\EndOfBibitem
\bibitem[Ghigo \latin{et~al.}(2004)Ghigo, Roos, and Malmqvist]{Gio04}
Ghigo,~G.; Roos,~B.~O.; Malmqvist,~P.-{\AA}. A Modified Definition of the
  Zeroth-Order Hamiltonian in Multiconfigurational Perturbation Theory
  (CASPT2). \emph{Chem. Phys. Lett.} \textbf{2004}, \emph{396}, 142--149\relax
\mciteBstWouldAddEndPuncttrue
\mciteSetBstMidEndSepPunct{\mcitedefaultmidpunct}
{\mcitedefaultendpunct}{\mcitedefaultseppunct}\relax
\EndOfBibitem
\bibitem[Werner \latin{et~al.}(2012)Werner, Knowles, Knizia, Manby, and
  Sch{\"u}tz]{molpro}
Werner,~H.-J.; Knowles,~P.~J.; Knizia,~G.; Manby,~F.~R.; Sch{\"u}tz,~M. Molpro:
  a general-purpose quantum chemistry program package. \emph{Wiley
  Interdisciplinary Reviews: Computational Molecular Science} \textbf{2012},
  \emph{2}, 242--253\relax
\mciteBstWouldAddEndPuncttrue
\mciteSetBstMidEndSepPunct{\mcitedefaultmidpunct}
{\mcitedefaultendpunct}{\mcitedefaultseppunct}\relax
\EndOfBibitem
\bibitem[Werner \latin{et~al.}(2019)Werner, Knowles, Knizia, Manby,
  {Sch\"{u}tz}, Celani, Gy\"orffy, Kats, Korona, Lindh, Mitrushenkov, Rauhut,
  Shamasundar, Adler, Amos, Bennie, Bernhardsson, Berning, Cooper, Deegan,
  Dobbyn, Eckert, Goll, Hampel, Hesselmann, Hetzer, Hrenar, Jansen, K\"oppl,
  Lee, Liu, Lloyd, Ma, Mata, May, McNicholas, Meyer, {Miller III}, Mura,
  Nicklass, O'Neill, Palmieri, Peng, Pfl\"uger, Pitzer, Reiher, Shiozaki,
  Stoll, Stone, Tarroni, Thorsteinsson, Wang, and Welborn]{MolPro19}
Werner,~H.-J.; Knowles,~P.~J.; Knizia,~G.; Manby,~F.~R.; {Sch\"{u}tz},~M.;
  Celani,~P.; Gy\"orffy,~W.; Kats,~D.; Korona,~T.; Lindh,~R.; Mitrushenkov,~A.;
  Rauhut,~G.; Shamasundar,~K.~R.; Adler,~T.~B.; Amos,~R.~D.; Bennie,~S.~J.;
  Bernhardsson,~A.; Berning,~A.; Cooper,~D.~L.; Deegan,~M. J.~O.;
  Dobbyn,~A.~J.; Eckert,~F.; Goll,~E.; Hampel,~C.; Hesselmann,~A.; Hetzer,~G.;
  Hrenar,~T.; Jansen,~G.; K\"oppl,~C.; Lee,~S. J.~R.; Liu,~Y.; Lloyd,~A.~W.;
  Ma,~Q.; Mata,~R.~A.; May,~A.~J.; McNicholas,~S.~J.; Meyer,~W.; {Miller
  III},~T.~F.; Mura,~M.~E.; Nicklass,~A.; O'Neill,~D.~P.; Palmieri,~P.;
  Peng,~D.; Pfl\"uger,~K.; Pitzer,~R.; Reiher,~M.; Shiozaki,~T.; Stoll,~H.;
  Stone,~A.~J.; Tarroni,~R.; Thorsteinsson,~T.; Wang,~M.; Welborn,~M. MOLPRO,
  version 2019.2, a package of ab initio programs. 2019; see
  https://www.molpro.net\relax
\mciteBstWouldAddEndPuncttrue
\mciteSetBstMidEndSepPunct{\mcitedefaultmidpunct}
{\mcitedefaultendpunct}{\mcitedefaultseppunct}\relax
\EndOfBibitem
\bibitem[Balasubramani \latin{et~al.}(2020)Balasubramani, Chen, Coriani,
  Diedenhofen, Frank, Franzke, Furche, Grotjahn, Harding, H{\"a}ttig, Hellweg,
  Helmich-Paris, Holzer, Huniar, Kaupp, Marefat~Khah, Karbalaei~Khani,
  M{\"u}ller, Mack, Nguyen, Parker, Perlt, Rappoport, Reiter, Roy, R{\"u}ckert,
  Schmitz, Sierka, Tapavicza, Tew, van W{\"u}llen, Voora, Weigend,
  Wody{\'n}ski, and Yu]{Bal20}
Balasubramani,~S.~G.; Chen,~G.~P.; Coriani,~S.; Diedenhofen,~M.; Frank,~M.~S.;
  Franzke,~Y.~J.; Furche,~F.; Grotjahn,~R.; Harding,~M.~E.; H{\"a}ttig,~C.;
  Hellweg,~A.; Helmich-Paris,~B.; Holzer,~C.; Huniar,~U.; Kaupp,~M.;
  Marefat~Khah,~A.; Karbalaei~Khani,~S.; M{\"u}ller,~T.; Mack,~F.;
  Nguyen,~B.~D.; Parker,~S.~M.; Perlt,~E.; Rappoport,~D.; Reiter,~K.; Roy,~S.;
  R{\"u}ckert,~M.; Schmitz,~G.; Sierka,~M.; Tapavicza,~E.; Tew,~D.~P.; van
  W{\"u}llen,~C.; Voora,~V.~K.; Weigend,~F.; Wody{\'n}ski,~A.; Yu,~J.~M.
  TURBOMOLE: Modular Program Suite for ab initio Quantum-Chemical and
  Condensed-Matter Simulations. \emph{J. Chem. Phys.} \textbf{2020},
  \emph{152}, 184107\relax
\mciteBstWouldAddEndPuncttrue
\mciteSetBstMidEndSepPunct{\mcitedefaultmidpunct}
{\mcitedefaultendpunct}{\mcitedefaultseppunct}\relax
\EndOfBibitem
\bibitem[Tur()]{Turbomole}
TURBOMOLE V7.3 2018, a development of University of Karlsruhe and
  Forschungszentrum Karlsruhe GmbH, 1989-2007, TURBOMOLE GmbH, since 2007;
  available from {\tt http://www.turbomole.com} (accessed 13 June 2016).\relax
\mciteBstWouldAddEndPunctfalse
\mciteSetBstMidEndSepPunct{\mcitedefaultmidpunct}
{}{\mcitedefaultseppunct}\relax
\EndOfBibitem
\bibitem[Krylov and Gill(2013)Krylov, and Gill]{Kry13}
Krylov,~A.~I.; Gill,~P.~M. Q-Chem: an Engine for Innovation. \emph{WIREs
  Comput. Mol. Sci.} \textbf{2013}, \emph{3}, 317--326\relax
\mciteBstWouldAddEndPuncttrue
\mciteSetBstMidEndSepPunct{\mcitedefaultmidpunct}
{\mcitedefaultendpunct}{\mcitedefaultseppunct}\relax
\EndOfBibitem
\bibitem[Frisch \latin{et~al.}(2016)Frisch, Trucks, Schlegel, Scuseria, Robb,
  Cheeseman, Scalmani, Barone, Petersson, Nakatsuji, Li, Caricato, Marenich,
  Bloino, Janesko, Gomperts, Mennucci, Hratchian, Ortiz, Izmaylov, Sonnenberg,
  Williams-Young, Ding, Lipparini, Egidi, Goings, Peng, Petrone, Henderson,
  Ranasinghe, Zakrzewski, Gao, Rega, Zheng, Liang, Hada, Ehara, Toyota, Fukuda,
  Hasegawa, Ishida, Nakajima, Honda, Kitao, Nakai, Vreven, Throssell,
  Montgomery, Peralta, Ogliaro, Bearpark, Heyd, Brothers, Kudin, Staroverov,
  Keith, Kobayashi, Normand, Raghavachari, Rendell, Burant, Iyengar, Tomasi,
  Cossi, Millam, Klene, Adamo, Cammi, Ochterski, Martin, Morokuma, Farkas,
  Foresman, and Fox]{Gaussian16}
Frisch,~M.~J.; Trucks,~G.~W.; Schlegel,~H.~B.; Scuseria,~G.~E.; Robb,~M.~A.;
  Cheeseman,~J.~R.; Scalmani,~G.; Barone,~V.; Petersson,~G.~A.; Nakatsuji,~H.;
  Li,~X.; Caricato,~M.; Marenich,~A.~V.; Bloino,~J.; Janesko,~B.~G.;
  Gomperts,~R.; Mennucci,~B.; Hratchian,~H.~P.; Ortiz,~J.~V.; Izmaylov,~A.~F.;
  Sonnenberg,~J.~L.; Williams-Young,~D.; Ding,~F.; Lipparini,~F.; Egidi,~F.;
  Goings,~J.; Peng,~B.; Petrone,~A.; Henderson,~T.; Ranasinghe,~D.;
  Zakrzewski,~V.~G.; Gao,~J.; Rega,~N.; Zheng,~G.; Liang,~W.; Hada,~M.;
  Ehara,~M.; Toyota,~K.; Fukuda,~R.; Hasegawa,~J.; Ishida,~M.; Nakajima,~T.;
  Honda,~Y.; Kitao,~O.; Nakai,~H.; Vreven,~T.; Throssell,~K.;
  Montgomery,~J.~A.,~{Jr.}; Peralta,~J.~E.; Ogliaro,~F.; Bearpark,~M.~J.;
  Heyd,~J.~J.; Brothers,~E.~N.; Kudin,~K.~N.; Staroverov,~V.~N.; Keith,~T.~A.;
  Kobayashi,~R.; Normand,~J.; Raghavachari,~K.; Rendell,~A.~P.; Burant,~J.~C.;
  Iyengar,~S.~S.; Tomasi,~J.; Cossi,~M.; Millam,~J.~M.; Klene,~M.; Adamo,~C.;
  Cammi,~R.; Ochterski,~J.~W.; Martin,~R.~L.; Morokuma,~K.; Farkas,~O.;
  Foresman,~J.~B.; Fox,~D.~J. Gaussian 16 {R}evision {A}.03. 2016; Gaussian
  Inc. Wallingford CT\relax
\mciteBstWouldAddEndPuncttrue
\mciteSetBstMidEndSepPunct{\mcitedefaultmidpunct}
{\mcitedefaultendpunct}{\mcitedefaultseppunct}\relax
\EndOfBibitem
\bibitem[Becke(1993)]{Bec93}
Becke,~A.~D. Density-Functional Thermochemistry. 3. The Role of Exact Exchange.
  \emph{J. Chem. Phys.} \textbf{1993}, \emph{98}, 5648--5652\relax
\mciteBstWouldAddEndPuncttrue
\mciteSetBstMidEndSepPunct{\mcitedefaultmidpunct}
{\mcitedefaultendpunct}{\mcitedefaultseppunct}\relax
\EndOfBibitem
\bibitem[Stephens \latin{et~al.}(1994)Stephens, Devlin, Chabalowski, and
  Frisch]{Fri94}
Stephens,~P.~J.; Devlin,~F.~J.; Chabalowski,~C.~F.; Frisch,~M.~J. Ab Initio
  Calculation of Vibrational Absorption and Circular Dichroism Spectra Using
  Density Functional Force Fields. \emph{J. Phys. Chem.} \textbf{1994},
  \emph{98}, 11623--11627\relax
\mciteBstWouldAddEndPuncttrue
\mciteSetBstMidEndSepPunct{\mcitedefaultmidpunct}
{\mcitedefaultendpunct}{\mcitedefaultseppunct}\relax
\EndOfBibitem
\bibitem[Adamo and Barone(1999)Adamo, and Barone]{Ada99}
Adamo,~C.; Barone,~V. Toward Reliable Density Functional Methods Without
  Adjustable Parameters: the PBE0 Model. \emph{J. Chem. Phys.} \textbf{1999},
  \emph{110}, 6158--6170\relax
\mciteBstWouldAddEndPuncttrue
\mciteSetBstMidEndSepPunct{\mcitedefaultmidpunct}
{\mcitedefaultendpunct}{\mcitedefaultseppunct}\relax
\EndOfBibitem
\bibitem[Ernzerhof and Scuseria(1999)Ernzerhof, and Scuseria]{Erz99}
Ernzerhof,~M.; Scuseria,~G.~E. Assessment of the Perdew--Burke--Ernzerhof
  Exchange-Correlation Functional. \emph{J. Chem. Phys.} \textbf{1999},
  \emph{110}, 5029--5036\relax
\mciteBstWouldAddEndPuncttrue
\mciteSetBstMidEndSepPunct{\mcitedefaultmidpunct}
{\mcitedefaultendpunct}{\mcitedefaultseppunct}\relax
\EndOfBibitem
\bibitem[Zhao and Truhlar(2008)Zhao, and Truhlar]{Zha08b}
Zhao,~Y.; Truhlar,~D.~G. The M06 Suite of Density Functionals for Main Group
  Thermochemistry, Thermochemical Kinetics, Noncovalent Interactions, Excited
  States, and Transition Elements: Two New Functionals and Systematic Testing
  of Four M06-Class Functionals and 12 Other Functionals. \emph{Theor. Chem.
  Acc.} \textbf{2008}, \emph{120}, 215--241\relax
\mciteBstWouldAddEndPuncttrue
\mciteSetBstMidEndSepPunct{\mcitedefaultmidpunct}
{\mcitedefaultendpunct}{\mcitedefaultseppunct}\relax
\EndOfBibitem
\bibitem[Yanai \latin{et~al.}(2004)Yanai, Tew, and Handy]{Yan04}
Yanai,~T.; Tew,~D.~P.; Handy,~N.~C. A New Hybrid Exchange-Correlation
  Functional Using the Coulomb-Attenuating Method (CAM-B3LYP). \emph{Chem.
  Phys. Lett.} \textbf{2004}, \emph{393}, 51--56\relax
\mciteBstWouldAddEndPuncttrue
\mciteSetBstMidEndSepPunct{\mcitedefaultmidpunct}
{\mcitedefaultendpunct}{\mcitedefaultseppunct}\relax
\EndOfBibitem
\bibitem[Chai and Head-Gordon(2008)Chai, and Head-Gordon]{Cha08b}
Chai,~J.~D.; Head-Gordon,~M. Long-range Corrected Hybrid Density Functionals
  with Damped Atom--Atom Dispersion Corrections. \emph{Phys. Chem. Chem. Phys.}
  \textbf{2008}, \emph{10}, 6615--6620\relax
\mciteBstWouldAddEndPuncttrue
\mciteSetBstMidEndSepPunct{\mcitedefaultmidpunct}
{\mcitedefaultendpunct}{\mcitedefaultseppunct}\relax
\EndOfBibitem
\bibitem[Folkestad \latin{et~al.}(2020)Folkestad, Kj{\o}nstad, Myhre, Andersen,
  Balbi, Coriani, Giovannini, Goletto, Haugland, Hutcheson, H{\o}yvik, Moitra,
  Paul, Scavino, Skeidsvoll, Tveten, and Koch]{eT}
Folkestad,~S.~D.; Kj{\o}nstad,~E.~F.; Myhre,~R.~H.; Andersen,~J.~H.; Balbi,~A.;
  Coriani,~S.; Giovannini,~T.; Goletto,~L.; Haugland,~T.~S.; Hutcheson,~A.;
  H{\o}yvik,~I.-M.; Moitra,~T.; Paul,~A.~C.; Scavino,~M.; Skeidsvoll,~A.~S.;
  Tveten,~{\AA}.~H.; Koch,~H. eT 1.0: An Open Source Electronic Structure
  Program with Emphasis on Coupled Custer and Multilevel Methods. \emph{J.
  Chem. Phys.} \textbf{2020}, \emph{152}, 184103\relax
\mciteBstWouldAddEndPuncttrue
\mciteSetBstMidEndSepPunct{\mcitedefaultmidpunct}
{\mcitedefaultendpunct}{\mcitedefaultseppunct}\relax
\EndOfBibitem
\bibitem[Parrish \latin{et~al.}(2017)Parrish, Burns, Smith, Simmonett,
  DePrince, Hohenstein, Bozkaya, Sokolov, Di~Remigio, Richard, Gonthier, James,
  McAlexander, Kumar, Saitow, Wang, Pritchard, Verma, Schaefer, Patkowski,
  King, Valeev, Evangelista, Turney, Crawford, and Sherrill]{Psi4}
Parrish,~R.~M.; Burns,~L.~A.; Smith,~D. G.~A.; Simmonett,~A.~C.;
  DePrince,~A.~E.; Hohenstein,~E.~G.; Bozkaya,~U.; Sokolov,~A.~Y.;
  Di~Remigio,~R.; Richard,~R.~M.; Gonthier,~J.~F.; James,~A.~M.;
  McAlexander,~H.~R.; Kumar,~A.; Saitow,~M.; Wang,~X.; Pritchard,~B.~P.;
  Verma,~P.; Schaefer,~H.~F.; Patkowski,~K.; King,~R.~A.; Valeev,~E.~F.;
  Evangelista,~F.~A.; Turney,~J.~M.; Crawford,~T.~D.; Sherrill,~C.~D. Psi4 1.1:
  An Open-Source Electronic Structure Program Emphasizing Automation, Advanced
  Libraries, and Interoperability. \emph{J. Chem. Theory Comput.}
  \textbf{2017}, \emph{13}, 3185--3197\relax
\mciteBstWouldAddEndPuncttrue
\mciteSetBstMidEndSepPunct{\mcitedefaultmidpunct}
{\mcitedefaultendpunct}{\mcitedefaultseppunct}\relax
\EndOfBibitem
\bibitem[Oddershede \latin{et~al.}(1985)Oddershede, Gr{\=u}ner, and
  Diercksen]{Odd85}
Oddershede,~J.; Gr{\=u}ner,~N.~E.; Diercksen,~G.~H. Comparison Between Equation
  of Motion and Polarization Propagator Calculations. \emph{Chem. Phys.}
  \textbf{1985}, \emph{97}, 303--310\relax
\mciteBstWouldAddEndPuncttrue
\mciteSetBstMidEndSepPunct{\mcitedefaultmidpunct}
{\mcitedefaultendpunct}{\mcitedefaultseppunct}\relax
\EndOfBibitem
\bibitem[Ben-Shlomo and Kaldor(1990)Ben-Shlomo, and Kaldor]{Ben90}
Ben-Shlomo,~S.~B.; Kaldor,~U. N$_2$ Excitations Below 15 eV by the
  Multireference Coupled-Cluster Method. \emph{J. Chem. Phys.} \textbf{1990},
  \emph{92}, 3680--3682\relax
\mciteBstWouldAddEndPuncttrue
\mciteSetBstMidEndSepPunct{\mcitedefaultmidpunct}
{\mcitedefaultendpunct}{\mcitedefaultseppunct}\relax
\EndOfBibitem
\bibitem[Neugebauer \latin{et~al.}(2004)Neugebauer, Baerends, and
  Nooijen]{Neu04}
Neugebauer,~J.; Baerends,~E.~J.; Nooijen,~M. Vibronic Coupling and Double
  Excitations in Linear Response Time-Dependent Density Functional
  Calculations: Dipole-Allowed States of N$_2$. \emph{J. Chem. Phys.}
  \textbf{2004}, \emph{121}, 6155--6166\relax
\mciteBstWouldAddEndPuncttrue
\mciteSetBstMidEndSepPunct{\mcitedefaultmidpunct}
{\mcitedefaultendpunct}{\mcitedefaultseppunct}\relax
\EndOfBibitem
\bibitem[Loos and Jacquemin(2020)Loos, and Jacquemin]{Loo20b}
Loos,~P.-F.; Jacquemin,~D. Is ADC(3) as Accurate as CC3 for Valence and Rydberg
  Transition Energies? \emph{J. Phys. Chem. Lett.} \textbf{2020}, \emph{11},
  974--980\relax
\mciteBstWouldAddEndPuncttrue
\mciteSetBstMidEndSepPunct{\mcitedefaultmidpunct}
{\mcitedefaultendpunct}{\mcitedefaultseppunct}\relax
\EndOfBibitem
\bibitem[Casanova-P{\'a}ez \latin{et~al.}(2019)Casanova-P{\'a}ez, Dardis, and
  Goerigk]{Cas19}
Casanova-P{\'a}ez,~M.; Dardis,~M.~B.; Goerigk,~L. $\omega$B2PLYP and
  $\omega$B2GPPLYP: The First Two Double-Hybrid Density Functionals with
  Long-Range Correction Optimized for Excitation Energies. \emph{J. Chem.
  Theory Comput.} \textbf{2019}, \emph{15}, 4735--4744\relax
\mciteBstWouldAddEndPuncttrue
\mciteSetBstMidEndSepPunct{\mcitedefaultmidpunct}
{\mcitedefaultendpunct}{\mcitedefaultseppunct}\relax
\EndOfBibitem
\bibitem[Peach \latin{et~al.}(2008)Peach, Benfield, Helgaker, and Tozer]{Pea08}
Peach,~M. J.~G.; Benfield,~P.; Helgaker,~T.; Tozer,~D.~J. Excitation Energies
  in Density Functional Theory: an Evaluation and a Diagnostic Test. \emph{J.
  Chem. Phys.} \textbf{2008}, \emph{128}, 044118\relax
\mciteBstWouldAddEndPuncttrue
\mciteSetBstMidEndSepPunct{\mcitedefaultmidpunct}
{\mcitedefaultendpunct}{\mcitedefaultseppunct}\relax
\EndOfBibitem
\bibitem[K{\'a}nn{\'a}r \latin{et~al.}(2017)K{\'a}nn{\'a}r, Tajti, and
  Szalay]{Kan17}
K{\'a}nn{\'a}r,~D.; Tajti,~A.; Szalay,~P.~G. Accuracy of Coupled Cluster
  Excitation Energies in Diffuse Basis Sets. \emph{J. Chem. Theory Comput.}
  \textbf{2017}, \emph{13}, 202--209\relax
\mciteBstWouldAddEndPuncttrue
\mciteSetBstMidEndSepPunct{\mcitedefaultmidpunct}
{\mcitedefaultendpunct}{\mcitedefaultseppunct}\relax
\EndOfBibitem
\bibitem[Casida \latin{et~al.}(1998)Casida, Jamorski, Casida, and
  Salahub]{Cas98}
Casida,~M.~E.; Jamorski,~C.; Casida,~K.~C.; Salahub,~D.~R. Molecular Excitation
  Energies to High-Lying Bound States from Time-Dependent Density-Functional
  Response Theory: Characterization and Correction of the Time-Dependent Local
  Density Approximation Ionization Threshold. \emph{J. Chem. Phys.}
  \textbf{1998}, \emph{108}, 4439--4449\relax
\mciteBstWouldAddEndPuncttrue
\mciteSetBstMidEndSepPunct{\mcitedefaultmidpunct}
{\mcitedefaultendpunct}{\mcitedefaultseppunct}\relax
\EndOfBibitem
\bibitem[Tozer and Handy(1998)Tozer, and Handy]{Toz98}
Tozer,~D.~J.; Handy,~N.~C. Improving Virtual {{Kohn}}\textendash{{Sham}}
  Orbitals and Eigenvalues: {{Application}} to Excitation Energies and Static
  Polarizabilities. \emph{J. Chem. Phys.} \textbf{1998}, \emph{109},
  10180--10189\relax
\mciteBstWouldAddEndPuncttrue
\mciteSetBstMidEndSepPunct{\mcitedefaultmidpunct}
{\mcitedefaultendpunct}{\mcitedefaultseppunct}\relax
\EndOfBibitem
\end{mcitethebibliography}

\end{document}